\documentclass[%
aip,
amsmath,amssymb,
reprint,nofootinbib%
]{revtex4-2}
\usepackage{graphicx,array,booktabs,rotating}% Include figure files
\usepackage{dcolumn}% Align table columns on decimal point
\usepackage{bm}% bold math
\usepackage{mathptmx}
\usepackage{multirow}
\usepackage{afterpage,float,color,xcolor}
\usepackage{hyperref}% add hypertext capabilities
\usepackage{etoolbox} % for \appto
\usepackage{tabularx,tabulary}
\usepackage[capitalize]{cleveref}
\usepackage{times}
\hypersetup{
	colorlinks,
	linkcolor={blue!100!black},
	citecolor={blue!100!black},
	urlcolor={blue!100!black}
}
\newcommand{\PRLsep}{\noindent\makebox[\linewidth]{\resizebox{0.3333\linewidth}{1pt}{$\bullet$}}\bigskip}
\makeatletter
\newcommand\footnoteref[1]{\protected@xdef\@thefnmark{\ref{#1}}\@footnotemark}
\makeatother

\newcommand{\etal}{\textit{et al.}}
\usepackage[page]{totalcount}
\usepackage{etoolbox,fancyhdr,xcolor}%
\usepackage[displaymath,pagewise]{lineno}% Enable numbering of text and display math
\begin{document}
\preprint{AIP/123-QED}	
	\title[\textbf{Journal to be decided} (2021) $|$  Manuscript prepared]{Unsteady dynamics in a subsonic duct flow with a bluff body}

	\author{Luckachan. K. George}
	%\email{luckachankgeorge@gmail.com}
	\affiliation{Department of Aerospace Engineering, Amrita School of Engineering, Coimbatore, Amrita Vishwa Vidyapeetham, 641112, India}%
	
	\author{S. K. Karthick}
	\email{skkarthick@ymail.com (Corresponding Author)}
	\affiliation{Faculty of Aerospace Engineering, Technion-Israel Institute of Technology, Haifa-3200003, Israel}%
	
	\author{A. R. Srikrishnan}
	%\email{ar_srikrishnan@cb.amrita.edu}
	\affiliation{Department of Aerospace Engineering, Amrita School of Engineering, Coimbatore, Amrita Vishwa Vidyapeetham, 641112, India}%
	
	\author{R. Kannan}
	%\email{r_kannan@cb.amrita.edu}
	\affiliation{Department of Aerospace Engineering, Amrita School of Engineering, Coimbatore, Amrita Vishwa Vidyapeetham, 641112, India}%
	
	\date{\today}% It is always \today, today,

\begin{abstract}
A series of reduced-order numerical simulations on a specific bluff body type (v-gutters) in a subsonic duct flow is done to assess the unsteady wake dynamics. Two of the v-gutter's geometrical parameters are varied: the v-gutter's base angle ($\theta$) and the size of a slit ($\xi$) at the leading-edge of the v-gutter. Turbulent flow kinematics and pressure field are analyzed to evaluate the unsteadiness at a freestream Mach number of $M_\infty = 0.25$ and a freestream Reynolds number based on bluff body's transverse length (L) of $Re_L=0.1 \times 10^6$. Five v-gutter angles are considered ($\theta$, $^\circ = \pi/6, \pi/4, \pi/3, 5\pi/12, \pi/2$) and three slit sizes ($\xi$, mm =0,0.25,0.5) are considered only for a particular $\theta = [\pi/6]$. In general, high fluctuations in velocity and pressure are seen for the bluffest body in consideration ($\theta = \pi/2$) with higher drag ($c_d$) and total pressure loss ($\Delta p_0$). However, bluffer bodies produce periodic shedding structures that promote flow mixing. On the other hand, the presence of a slit on a streamlined body ($\theta = \pi/6$) tends to efficiently stabilize the wake and thus, producing almost a periodic shedding structure with lower $c_d$ and $\Delta p_0$. For $\theta = [\pi/6]$, broadened spectra in vortex shedding is seen with a peak at $[fL/u_\infty] \sim 0.08$. For $\theta \geq [\pi/4]$, a dominant discrete shedding frequency is seen with a gradual spectral decay. Similarly, the effects of $\xi$ on the $\theta = [\pi/6]$ case produce a discrete shedding frequency instead of a broadened one, as told before. The shedding frequency increases to a maximum of $[fL/u_\infty] \sim 0.26$ for the maximum slit size of $\xi = 0.5$. From the analysis of the $x-t$ diagram and the modal analysis of vorticity and velocity magnitude in the wake, the peaks are indeed found to agree with the spectral analysis. More insights on the shedding vortices, momentum deficit in the wake, varying energy contents in the flow field, and the dominant spatiotemporal structures are also provided. 
\end{abstract}
 

\maketitle

\section{Introduction}\label{sec:introduction}
The study of fluid flow characteristics downstream of the bluff bodies lately garnered a lot of interest due to the relevance in designing a variety of duct flow systems \cite{mair_1971,perry_1982,hunt_1973,balachandar_1997}, especially combustors. Such a study helps promote the energy transfer between the wake zone and the flow field surrounding the bluff body. The findings are effectively utilized in flame stabilization methods, mainly to improve the combustor and afterburner performance in ramjets and gas turbine engines\cite{Wei2021,Balasubramaniyan2021,Tayyab2021}. If not appropriately sustained, the flame in the combustion chamber encounters blowout phenomena due to the high velocity unburnt air-fuel mixture. Such an event reduces the combustion efficiency owing to fuel spillage. Flame stabilization in the aerodynamic wake of a bluff body is one of the passive techniques that come in handy. However, even for a cold (non-reacting) flow, the presence of a bluff body in a ducted flow offers hindrance in terms of dominant acoustic loading on the duct due to wake induced unsteadiness, higher drag coefficient ($c_d$), and severe total pressure loss ($\Delta p_0$). Event of combustion and flame holding further complicated the fluid dynamics in the duct. Such adverse outcomes motivated researchers to explore multiple flame stabilization methods to prevent blowout behavior at high velocities. 

On a general note, for a hot (reacting) flow process in a duct, the thermal diffusivity, blockage ratio, laminar flame speed, and aerodynamic wake width are the primary factors that influence the flame blowout  \cite{pitts_1989,williams_1949,chao_2000,potter_1958}. Extensive studies on flame stabilization strategies confirm bluff body as the most commonly adopted technique for combustion applications, as it provides effective flame retention with comparatively minimum loss in $\Delta p_0$ \cite{fan_2014,blanchard_2014,umyshev_2017,el_1994,li_2016,km_}. One of the popular bluff body shapes is the v-gutter\cite{Bush2007,yang_1992,yang_1994,George_2021} which outperforms all the other bluff body shapes in terms of energy and mass transfer of both unburnt air-fuel mixture and high-energy combustion particles. The wake flow structure behind the v-gutter plays an important role in flame stabilization. It ensures mass transfer in and out of the recirculation zone, which aids in establishing sustained combustion. The unburnt mixture is transported into the wake zone by the continual shedding of vortices. The high-energy combusted products transfer the energy to the remaining mixture in the combustors. Thus, the required flow field for flame stabilization pertains only to the wake characteristics of a bluff body; it can even be studied in cold flow for simplicity and safety.

Multiple studies have examined the hot and cold flow behaviour over bluff bodies to assess flame stabilisation effectiveness\cite{yue_2003,zukoski_1955,prasad_1997,fujii_1981,longwell_1948}. Perry \etal\cite{perry_1982} used flow visualization techniques to study the vortex shedding characteristics behind the circular cylinder and captured the streamline pattern behind the bluff body. King\cite{king_1957} experimentally investigated the combustion and flow field characteristics behind the v-gutter bluff body. The author suggested an equation for a flame blowout at high velocities, represented by the ratio between the residence time of the unburnt mixture and the chemical ignition time. Smith \etal\cite{smith_2007} numerically and experimentally investigated the reacting and non-reacting flow behind the v-gutter and concluded numerical analysis results to be matching with experimental results in terms of blowout velocity. The authors also explored different vortex shedding characteristics like von Karman vortex street behind the v-gutter. Kahawita and Wang\cite{kahawita_2002} performed numerical simulation over trapezoidal bluff bodies and analyzed the effect of trapezoidal shape on vortex shedding. The authors concluded that the bluff body size and shape play an important role in the wake structure formation in the downstream profile. In addition, they observed that the Strouhal number ($fL/u_\infty$) and vortex shedding pattern are dramatically affected when the trapezoidal height ($L$) rises. Vortex shedding is one of the basic techniques to achieve mixing enhancement \cite{gillies_1998}. However, it imposes design criticality on the bluff body as it begins to experience varying lift coefficient ($c_l$) fluctuations and mean drag coefficient ($c_d$) variations \cite{karthikeyan_2017}. 

The study of induced or shedding vortices by the bluff body is also useful in a variety of other aerospace applications, such as noise reduction in exhaust nozzles\cite{subramanian_2018} and liquid drain in fuel tanks\cite{ajith_2016,ajith_2017,prabhu_2021}. The induced vortices over the different bluff body shapes have also been investigated by some of the researchers computationally and experimentally to understand the unburnt species mixing\cite{bruno_20103,cai_2008} as well as the stability of the flame\cite{li_2016large}. Kiya \etal\cite{kiya_2001} reviewed the wake flow behind different shaped bluff bodies like a sphere, circular disks, and elliptical disks and concluded that the frequency ($f$) of alternative shedding crucially depends on Reynolds number ($Re_L$). Bruno \etal\cite{bruno_20103} performed a numerical investigation of flow over two and three-dimensional rectangular bluff bodies at high $Re_L$. Karthikeyan \etal\cite{karthikeyan_2019} numerically simulated and analyzed the combined effects of vortex shedding, undulation, and jet-wake dominant phenomena over a circular cylinder. The authors identified properties such as $c_l$, $c_d$ and power spectral density (PSD). Briones and Sekar\cite{briones_2012} also studied the fluid flow behavior and vortex shedding patterns of various bluff body geometries with an emphasis on $c_d$.  

In the early decades, some of the researchers\cite{yang_1992,yang_1994} showed that a perforated bluff body shows better results in terms of stagnation pressure losses instead of using a solid v-gutter bluff body. Yang and Tsai\cite{yang_1992} experimentally investigated the flow recirculation and vortex shedding characteristics behind the open slit v-gutter. They observed an asymmetrical wake structure developing behind the bluff body due to the flow penetration through open slits. Similarly, Yang \etal\cite{yang_1994} performed flame stabilization behind the open slit v-gutter by identifying the parameters like velocity, pressure, turbulent intensity, and temperature characteristics. They observed an improvement in turbulence in the downstream section due to open slits.

Nishino\etal\cite{nishino_2008} performed unsteady Reynolds-averaged Navier Stokes (URANS) and detached-eddy simulations (DES) behind the circular bluff body and analyzed the characteristics of induced wake developed behind the bluff body. The wake dynamics behind the bluff body are well predicted by DES simulations and are in good accord with experimental vortex shedding characteristics. Breuer\cite{breuer_1998,michael_2000} used large-eddy simulations (LES) on a circular cylinder and came up with some interesting findings. Travel \etal\cite{travin_2000} performed numerical simulations over a circular cylinder and captured the features of vortex shedding and pressure coefficients using the LES model. The findings are shown to be in good agreement with the experiment.

Based on the available literature on v-gutter type bluff bodies, one can conclude that there is an inadequacy in explaining the effect of v-gutter's geometrical parameters like the angle ($\theta$) and slit sizes ($\xi$) on the critical flow physics like the unsteady wake, mean flow properties, acoustic loading, forces on the v-gutters, the total pressure loss in the duct, and dominant flow structures responsible for fluid mixing. Hence, the authors are motivated by the reasons mentioned above to formulate the following distinct objectives:
\begin{enumerate}
\item {To specifically begin a two-dimensional computational campaign using DES type of simulation towards understanding the effect of v-gutter's span angle on wake-vortex shedding since it has the potential to affect the recirculation zone where the mass/momentum transfer occurs.}
\item {To study the effect of v-gutter central slit size on the wake-vortex morphology, as the slit has the potential to affect the width and strength of the wake zone by allowing some amount of freestream fluid mass to penetrate the v-gutter's wake.}
\item {To understand the unsteady flow dynamics in the wake using spectral analysis, monitoring the force fluctuations, and quantifying the acoustic loading on the duct itself as these components tend to invoke fluid-structure interactions, which are inherently detrimental.}
\item {To identify the dominant Spatio-temporal modes responsible for the distinct wake characteristics for each of the cases in consideration: v-gutter angle and slit-size, using modern data analytic tools like the modal analysis, in particular, Proper Orthogonal Decomposition (POD) for extracting energetic spatial modes, and Dynamic Mode Decomposition (DMD) for identifying the dynamic temporal modes.}
\end{enumerate}

Asides the introduction on the particular research topic in Sec. \ref{sec:introduction}, the rest of the manuscript is organized as follows. A brief description on the problem statement and the flow conditions are given in Sec. \ref{sec:prob_state}. Details on the numerical methodology, including the boundary conditions, grid and time independence, and solver validation are provided in Sec. \ref{sec:num_meth}. Results and discussions after analysing the data from the computational solvers are elaborately given in different subsections at Sec. \ref{sec:res_disc}. Following are some of the broad subsection under which the results and discussions are made: instantaneous (Sec. \ref{ssec:inst_flow}) and time averaged (Sec. \ref{ssec:time_avg}) flow field, total pressure loss (Sec. \ref{ssec:p_loss}), unsteady dynamics (Sec. \ref{ssec:unsteady_flow}), momentum deficit (Sec. \ref{ssec:momentum}), and spatio temporal modes (Sec. \ref{ssec:spatiotemporal}). All the vital results are listed briefly in the conclusions section at Sec. \ref{sec:conclusion}.

\section{Problem Statement}\label{sec:prob_state}
\begin{figure*}
	\centering{\includegraphics[width=\textwidth]{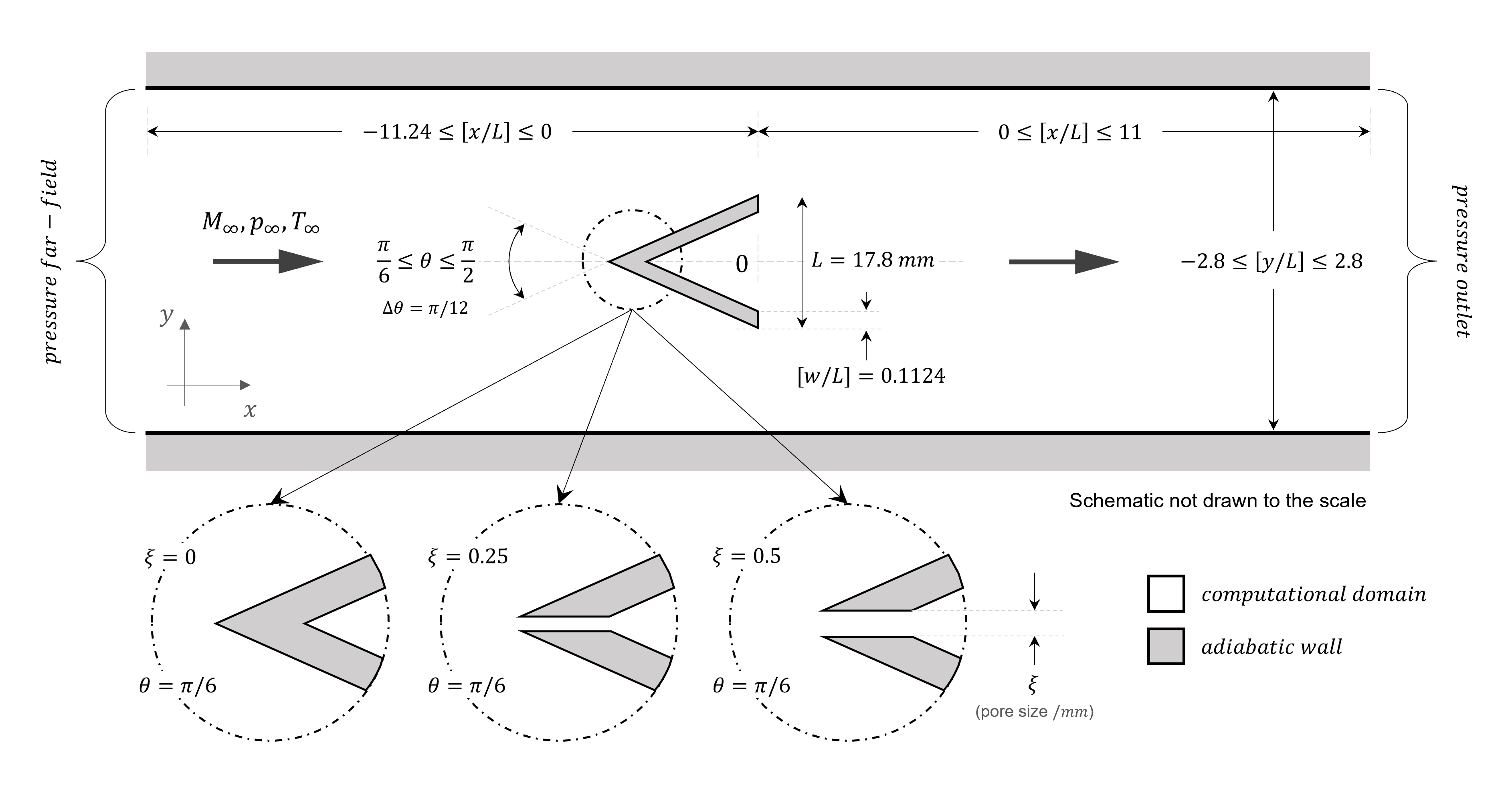}}
	\caption{Schematic representation of the problem statement highlighting v-gutter's different geometrical configurations (variation of v-gutter's angle at $\xi=0$ mm: $\theta(^\circ)=[\pi/6,\pi/4,\pi/3,5\pi/12,\pi/2]$ with $\Delta \theta = [\pi/12]$ and variation of v-gutter's slit size at $\theta=[\pi/6]$: $\xi=[0, 0.25,0.5]$ in mm with $\Delta \xi = 0.25$ mm), the axes system, the origin, the computational domain limits and the boundary conditions. Flow is from the left to the right. The v-gutter's width (shoulder tip-to-tip) is kept constant at $L=17.8$ mm and it is used for non-dimensionalizing the length parameters. The schematic is not drawn to the scale.}
	\label{fig:schematics}
\end{figure*}

The authors consider a simplistic v-gutter geometry as a bluff body that is immersed to the center in a two-dimensional (2D) wall bounded flow ($xy$-plane). The 2D duct height and length are considered in the limits of $-2.8\leq [y/L] \leq 2.8$ and $-11.3 \leq [x/L] \leq 11$, respectively. The origin is at the middle of the v-gutter where the tip-to-tip length ($L=17.8$ mm) of the v-gutter is measured. The origin demarcates the flow zones as upstream ($[x/L]<0$) and downstream ($[x/L]>0$). All the length related quantities in the upcoming sections are non-dimensionalized with respect to $L$. The v-gutter has a right-opening angle $\theta (^\circ)$ towards the downstream that is varied as $[\pi/6] \leq \theta \leq [\pi/2]$ in steps of $\Delta \theta = [\pi/12]$ for $\xi=0$. Similarly, the slit size along the v-gutter's center is changed in three steps: $\xi = [0,0.25,0.5]$ for $\theta = [\pi/6]$, where $\xi$ is given by the local slit size in mm. A typical schematic of the computational flow problem is given in Figure \ref{fig:schematics}.   

The flow conditions are selected after carefully considering the range of flow velocities expected in the combustor of a typical ramjet or the afterburner of a turbojet engine \cite{mattingly_2016}. However, only cold-flow (non-reacting) conditions are considered prominently in the open literature, just like in the experiments \cite{yang_1992} and computations \cite{blanchard_2014} performed on similar types of flows towards understanding the underlying flow physics. Hence, the authors have adopted similar flow conditions in the present studies. A typical freestream velocity of $u_\infty = 85$ m/s is considered and a freestream Reynolds number of $Re_L=0.1 \times 10^6$ is achieved in the test-section based on the v-gutter's tip-to-tip length ($L$). The other thermodynamic variables correspond to the sea-level operating conditions, and a freestream Mach number of $M_\infty=0.25$ is obtained while using air as the ideal gas. A typical freestream flow condition achieved in the computational domain's inlet is tabulated in Table \ref{table:flow_cond}.

\begin{table}
	\caption{Flow conditions achieved in the inlet based on the sea-level standard conditions for the considered v-gutter problem statement under different geometrical configurations\footnote{Two geometrical parameters are varied: 1. different v-gutter angles represented as $\theta$ in degree for $\xi=0$: $\theta=[\pi/6,\pi/4,\pi/3,5\pi/12,\pi/2]$ with $\Delta \theta = [\pi/12]$, 2. different slit sizes as $\xi$ in mm for $\theta=[\pi/6]$: $\xi=[0, 0.25,0.5]$ with $\Delta \xi=0.25$}.}
		\label{table:flow_cond}
		\begin{ruledtabular}
			\begin{tabular}{@{}ll@{}}    
				\textbf{Quantities} &
				\textbf{Values}\\ 
				\midrule
				Specific heat ratio ($\gamma$, air as ideal gas) & 1.4\\
				Total Pressure ($p_0\times 10^5$, Pa) &
				1.08 \\
				Total Temperature ($T_0$, K) &
				293.3\\
				Freestream Temperature ($T_\infty $, K) &
				288\\
				Freestream Pressure ($p_\infty \times 10^5$, Pa) &
				1.01\\
				Freestream Velocity ($u_\infty$, m/s) &
				85\\
				Freestream Kinematic Viscosity ($\nu_\infty\times 10^{-5}$, m$^2$/s) &
				1.46  \\
				Freestream Density ($\rho_\infty \times 10^{-3}$, kg/m$^{3}$) &
				1.23\\
				Freestream Mach number ($M_\infty$) &
				0.25 \\
				Reynolds number\footnote{$Re_L=[u_\infty L/\nu_\infty]$, where the v-gutter's width is kept constant and it is taken as the reference length of $L=17.8$ mm} ($Re_L \times 10^{6}$)  &
				0.1 \\    			 
			\end{tabular}
		\end{ruledtabular}
	\end{table}

\section{Numerical Methodology} \label{sec:num_meth}
A commercial solver is used to resolve the fundamental equations of fluid motion in a two-dimensional sense. Ansys-Fluent\textsuperscript{\tiny\textregistered}\cite{Fluent_2013} is known to resolve these equations well for problems like the separated and wake flows at moderate flow velocities\cite{Bai2018,Fukuda2017,Zhou2019,Li2021}. Recently, the Detached Eddy Simulation (DES) module in the considered flow solver has been adopted popularly\cite{Pandey2022,Chode2021} to study v-gutter type time-dependent wake-vortex flow problems. The flow field of the v-gutter is a classic bluff body problem, where the wake characteristics are periodic with turbulent coherent structures present downstream. Although two-dimensional solutions are not practical, most of the flow features and the outcomes, like the first dominant spatiotemporal oscillation mode, are the same between 2D and 3D. In fact, in the upcoming sections, it is shown that the variations coming from the lateral modes are less and dominant features are indeed identifiable with the two-dimensional solutions.

\subsection{Grid generation, and solver description} \label{ssec:mesh_solver}

\begin{figure*}
	\centering{\includegraphics[width=0.9\textwidth]{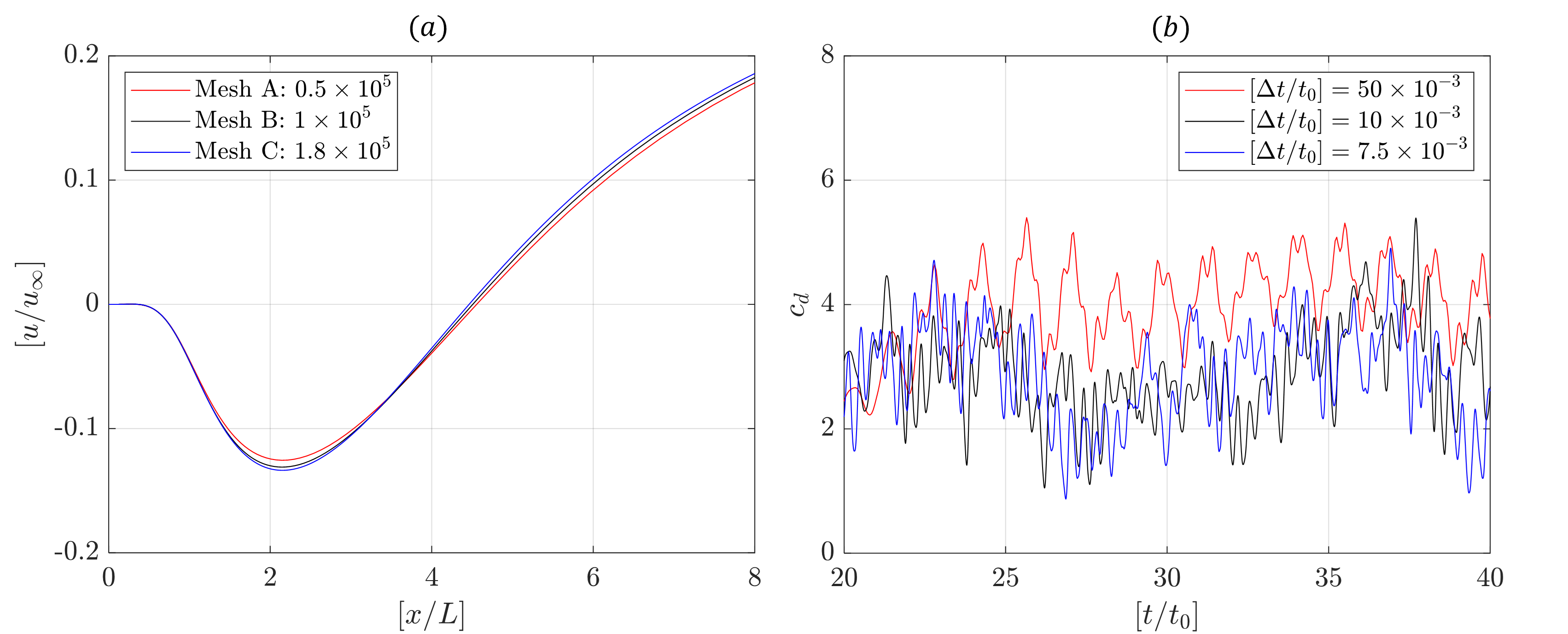}}
	\caption{The computational case of Blanchard \etal \cite{blanchard_2014} is considered to prepare and assess the mesh counts and solver time-step. (a) Graph showing the computed dimensionless streamwise velocity ($u/u_\infty$) variation along the centerline ($[y/L]=0$) from the RANS based simulation for three different mesh densities (Mesh A,B and C with 0.5, 1.0, and 1.8 $\times 10^5$ mesh cell counts). (b) Graph showing the DES based time-step independence study by monitoring the drag co-efficient ($c_d$) variation of the v-gutter against three different dimensionless time-steps ($[\Delta t/t_0] \times 10^{-3}$, where $t_0=1$ ms): 50 (coarse), 10 (medium) and 7.5 (fine).}
	\label{fig:grid_time_independence}
\end{figure*}

Firstly, Reynolds Averaged Navier-Stokes (RANS) equations without unsteady terms are solved, and then, the solution is sought with unsteady terms (URANS). Later, the Detached Eddy Simulation (DES) begins, and the solution is allowed to run for a few time steps. A stationary oscillation of the unsteady terms, for example the drag co-efficient ($c_d=D/[2\rho_\infty u_\infty^2L]$, where $D$ is the total drag per m, N/m), is achieved after $[t/t_0>20]$ and the solution is ran for $20\leq [t/t_0]\leq 40$. At least 10 to 15 oscillation cycles are completely captured during the stationary solution acquisition. All the unsteady statistics are taken during the stationary solution time and are discussed further in the subsequent sections. 

Meshing is performed using Ansys-ICEM\textsuperscript{\tiny\textregistered}\cite{Icem_2013} module. A two-dimensional grid is generated with quadrilateral cells stacked in a structured manner. Among the total number of cells, nearly 95\% of the cells have an equisize skewness parameter of 0.2. A first cell distance of 10 $\mu$m is maintained in every wall boundary, ensuring the turbulence wall parameter is $y^+\leq1$. Nowhere downstream the v-gutter, the progression of cell spacing is increased by more than 1.2 to resolve the wake vortices properly. The cell counts in the downstream regions are adjusted to develop a mesh of different densities whose influence will be discussed in the upcoming section. The top and bottom edges of the computational domain are constrained by the no-slip boundary condition and kept at adiabatic conditions. The front and rear edges of the numerical domain are applied with pressure far-field and pressure outlet boundary conditions.  

A pressure-based coupled solver is sought after to resolve the wake-vortex dynamics. As discussed in the upcoming section, a steady-state solution methodology is adopted for sizing the mesh counts. Asides, transient schemes are considered for the rest. Air as ideal gas is chosen as the working fluid, and hence, the energy equation is also solved in parallel with the continuity and momentum equations. Although modeling viscosity terms is less critical due to the flow field's less compressibility, dynamic viscosity is modeled using Sutherland's three coefficient method. Near-wall turbulence is modeled through RANS equations with vorticity-strain-based Spalart-Allmaras equations, whereas in the free shear layer and wakes, delayed detached eddy modeling is adopted. The constants for the modeling functions are kept default as given in the solver settings. 

A Green-Gauss node-based spatial discretization scheme is used to resolve the gradients. A second-order scheme is used to discretize the pressure terms, whereas a second-order upwind scheme is used to discretize the density, modified viscosity, and energy terms. Momentum terms are discretized using a bounded central difference scheme. A second-order implicit scheme is used for temporal discretization. Parameters like the explicit and under relaxation factors and the Courant numbers are kept as default values in the settings. Residuals from each iteration are scaled locally for all the fluid flow equations with the finalized mesh density and solver time-step as described in the next section. A residual criterion of at least $10^{-5}$ is achieved in the continuity, momentum, and energy equations for all the cases under consideration.

\subsection{Mesh size and time-step independence studies} \label{ssec:mesh_ind}

Three meshes are considered to assess the dependency of solutions on the mesh counts. Following mesh counts are generated to simulate different mesh densities: Mesh A (coarse): $0.5 \times 10^5$, Mesh B (medium): $1.0 \times 10^5$, and Mesh C (fine): $1.8 \times 10^5$. Case of Blanchard \etal \cite{blanchard_2014} is used as a benchmark for selecting an appropriate mesh density. The flow conditions are given as follows: $u_\infty= 35.5$ m/s, $L= 0.0305$ m, $T_\infty = 298$ K, and $Re_L\sim 0.08 \times 10^6$. The variation of dimensionless centerline ($[y/L]=0$) streamwise velocity ($u/u_\infty$) is monitored as shown in Figure \ref{fig:grid_time_independence}a for all the cases. A negligible variation of less than a percentage is seen in the distribution between the cases. Hence, to save the expense of computational power, a mesh with the course density is selected, which is Mesh A in this case.

\begin{figure*}
	\centering{\includegraphics[width=0.9\textwidth]{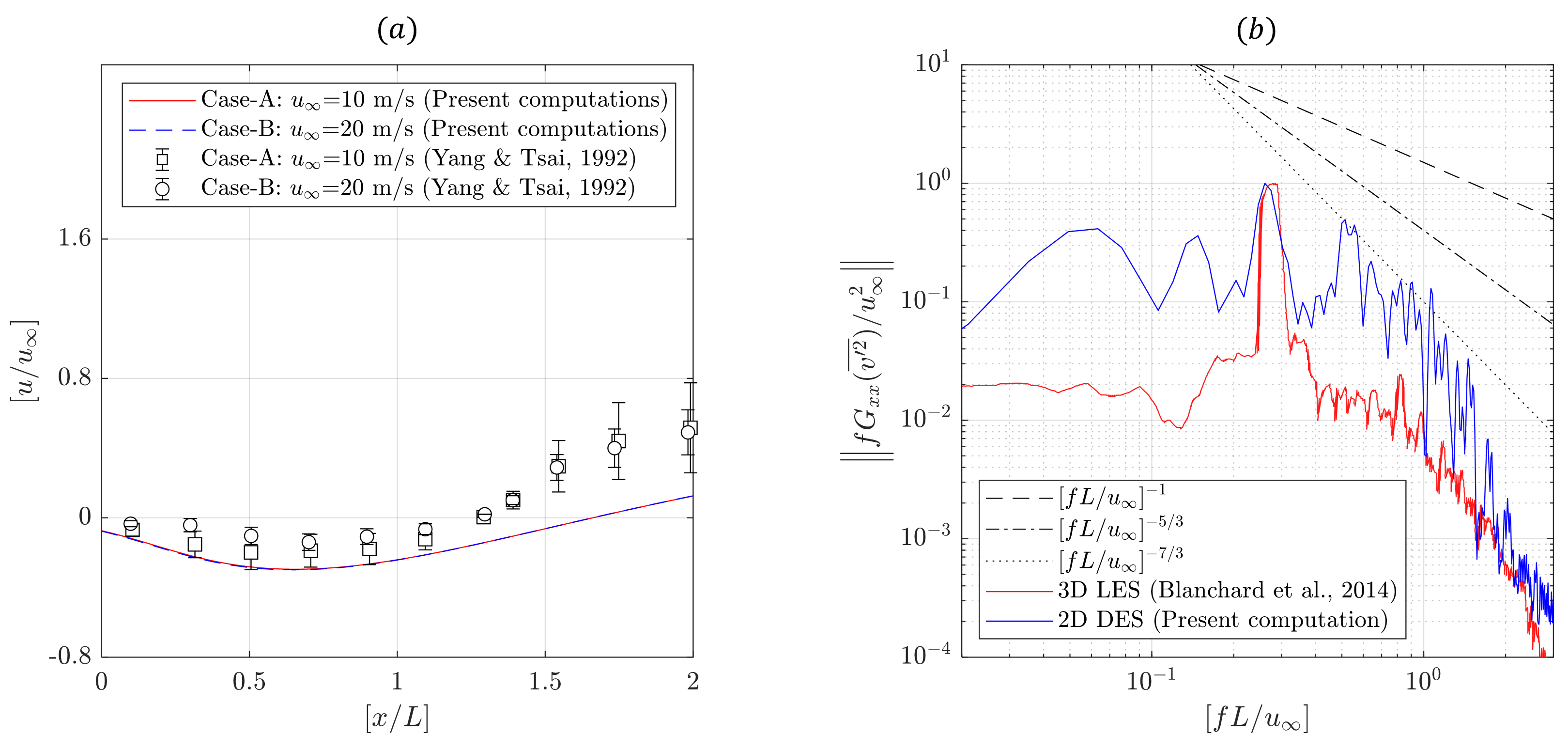}}
	\caption{(a) The experimental case of Yang and Tsai\cite{yang_1992} is considered for validating the steady state centerline ($y/L=0$) streamwise velocity decay ($u/u_\infty$). The graph shows the comparison of two different DES based computations (where $u_\infty=[10,20]$ m/s) with that of the experiments while using the Mesh A type cell counts. Actual error in the experiments are $\pm5\%$ of the mean value, however, for the sake of clarity, 10 times amplification is done. (b) Computational case of Blanchard \etal \cite{blanchard_2014} is taken to compare one of the dimensionless normalized Reynolds stress components' spectra $\left(\Vert fG_{xx}\overline{v'^2}/u_\infty^2 \Vert \right)$ with that of the present DES based computations at a point in the fluid domain defined by $[x/L,y/L]=[5,-0.5]$. The DES simulation is performed with a dimensionless time-step of $[\Delta t/t_0]=10\times 10^{-3}$ (where $t_0=1$ ms). The decaying spectra is compared against an arbitrary decay profile of -1. -5/3, and -7/3 to portray the play of turbulence scales.
}
	\label{fig:validation_steady_unsteady}
\end{figure*}

Three different time-steps are taken for the selected mesh which are represented in a dimensionless manner as follows: a. $[\Delta t/t_0]=50 \times 10^{-3}$ (coarse), b. $[\Delta t/t_0]=10 \times 10^{-3}$ (medium), and c. $[\Delta t/t_0]=7.5 \times 10^{-3}$ (fine), where $t_0 = 1$ ms. The same case of Blanchard \etal \cite{blanchard_2014} is considered for the time-independence study too. The v-gutter's drag coefficient ($c_d$) value is monitored during the stationary solution period for all the cases and plotted in Figure \ref{fig:grid_time_independence}b. An almost similar signal as ensured by the cross-correlation value which is larger than 0.9 is seen between the medium and fine time-steps. As in the previous case, to save the computing time, a medium time-step of $[\Delta t/t_0]=10 \times 10^{-3}$ is selected for running the further cases.

\subsection{Solver validation in spatial and spectral domain}

\begin{figure*}
	\centering{\includegraphics[width=0.9\textwidth]{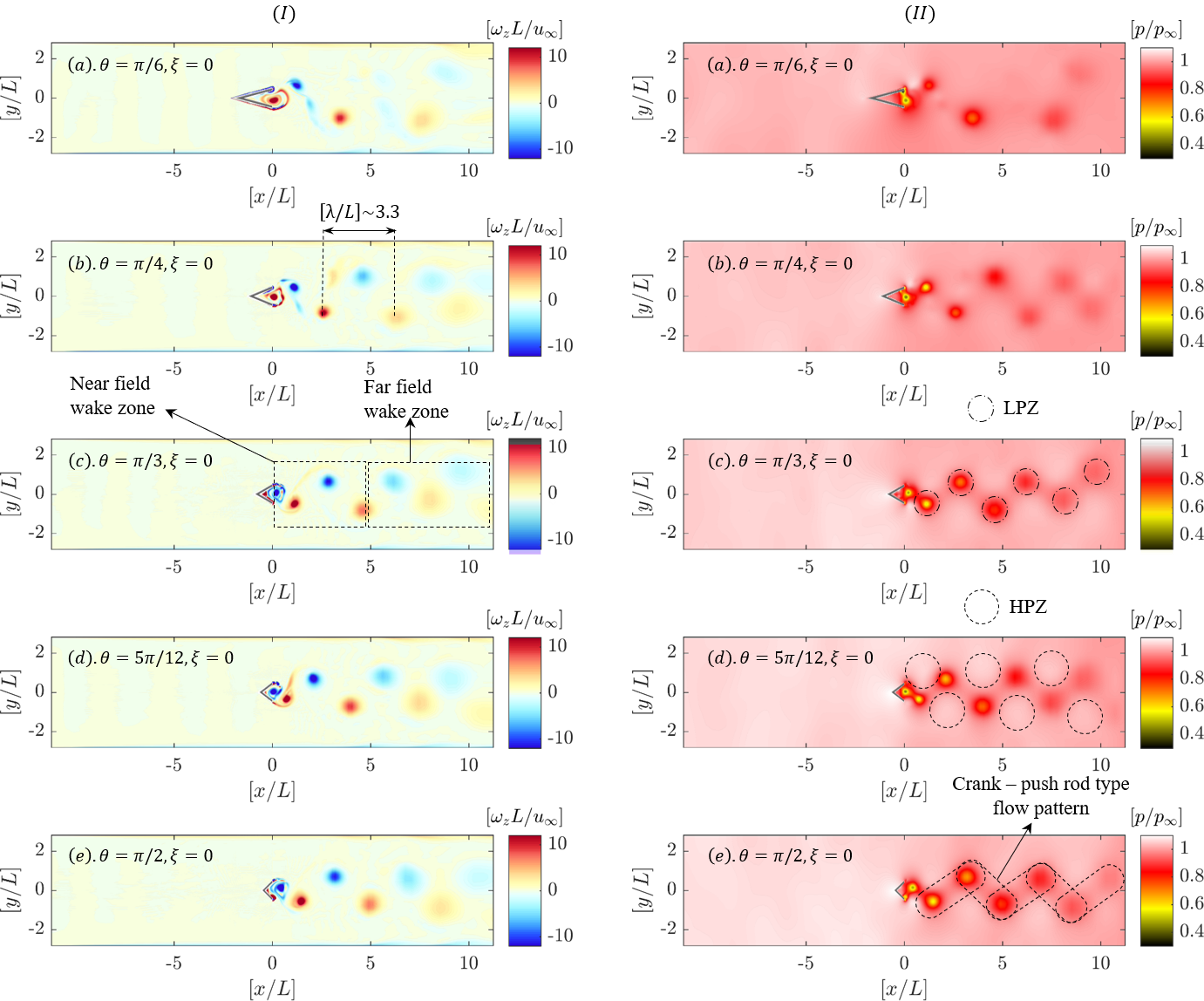}}
	\caption{\href{https://youtu.be/veqAQKOCUk0}{(Multimedia view)} Instantaneous contour plots observed at an arbitrary time step during the simulation for different v-gutter angles in degree: $\theta=[\pi/6,\pi/4,\pi/3,5\pi/12,\pi/2]$ at $\xi=0$ mm. (a) Non-dimensionlized vorticity about the z-axis $[\omega_{z}L/u_\infty]$ and (b) Non-dimensionlized  static pressure $[p/p_\infty]$. Vorticity is non-dimensionlized using v-gutters width of $L=17.8$ mm and freestream velocity of $u_\infty =85 m/s$. The static pressure is non-dimensionlized using freestream pressure of $p_\infty =1.01\times 10^5$, Pa.}
	\label{fig:inst_vor_press_angle}
\end{figure*}

Ansys-Fluent\textsuperscript{\tiny\textregistered} solver is prominently validated for steady-state and unsteady-state for a wide range of flow problems at varying speed regimes while solving in the RANS and URANS mode. However, as explained in the previous sections, bluff-body wake flows or separated flows, especially for two-dimensional cases, validations are scarcely available. Recently, in one of the authors' previous works, the capability of this particular solver is demonstrated in both URANS\cite{Sekar2020} and DES\cite{Karthick2021} modes at supersonic speed, where separation is encountered a lot. In line with those studies, validation is done for the bluff body flows like the v-gutters in a subsonic duct. Experimental cases of Yang and Tsai \cite{yang_1992} are considered for comparing the spatial variations of the centerline velocity with the present CFD from DES as shown in Figure \ref{fig:validation_steady_unsteady}a. The range of flow conditions achieved by Yang and Tsai are given as follows: $u_\infty= 1-20$ m/s, $L= 0.02$ m, and $Re_L\sim 0.12-2.4 \times 10^4$. A stationary unsteady flow for a run-time of $t=$5 ms is considered for collecting the turbulent statistics. After analysis, the centerline streamwise velocity data is compared with the experiments. Two test cases with a freestream velocity of $u_\infty=$ 10 and 20 m/s in the experiments are simulated in the CFD. The values obtained from the CFD almost match closely with each other. However, between the experiments, they begin to deviate noticeably in the downstream zones ($1.25 \leq [x/L]\leq 2$) and match fairly well within the uncertainty limits of both experiments and computations ($0 \leq [x/L]\leq 1.2$). 

Spectral measurements at any point in the fluid flow of a v-gutter are available in limited numbers. Hence, the computational results of Blanchard \etal \cite{blanchard_2014} are considered for the spectral validation. The authors validated the three-dimensional LES (Large Eddy Simulation, Fluent) solver with the experiments. However, the downstream spectra are given only from the computations. The authors took a point downstream the v-gutter at $[x/L,y/L]=[5,-0.5]$, which lies in the wake-vortex zone. Then, they compared one of the Reynolds stresses ($\sqrt{\overline{v'^2}}$) decay rate with the computations available in the literature as shown in Figure \ref{fig:validation_steady_unsteady}b. A dominant dimensionless shedding frequency of $[fL/u_\infty]\sim 0.25$ matches the present and previous computations. The decay rate also appears to be the same between the two cases, with a trend following $[fL/u_\infty]^{-5/3}$ at higher frequencies.

The following conclusions can be drawn from the spatial and temporal validation of a 2D solver: 1. The downstream deviations in the spatial data can be attributed to the three-dimensional geometrical effects arising from the wind-tunnel walls. Also, the physical flow instabilities in the lateral direction influence the wake vortex significantly in real-time; 2. Although the dominant spectra and the trend matches, the energy contents are slightly different. The energy contents at lower frequencies do not match the 2D (present) and 3D (literature) computations. The reason is attributed to the run-time considered to evaluate the spectra. In the literature, the flow time scales are not explicitly given, but they could be identified to be less than $t\lesssim$5 ms. The time-step also appeared to be $\Delta t=$100 $\mu$s. In contrast, the present simulation is done with a $\Delta t=$10 $\mu$s The amplification of fewer frequencies in the mid-frequency range is attributed to the absence of bounding walls and the constraints imposed on the growth of lateral instabilities.

\section{Results and discussions} \label{sec:res_disc}

\begin{figure*}
	\centering{\includegraphics[width=0.9\textwidth]{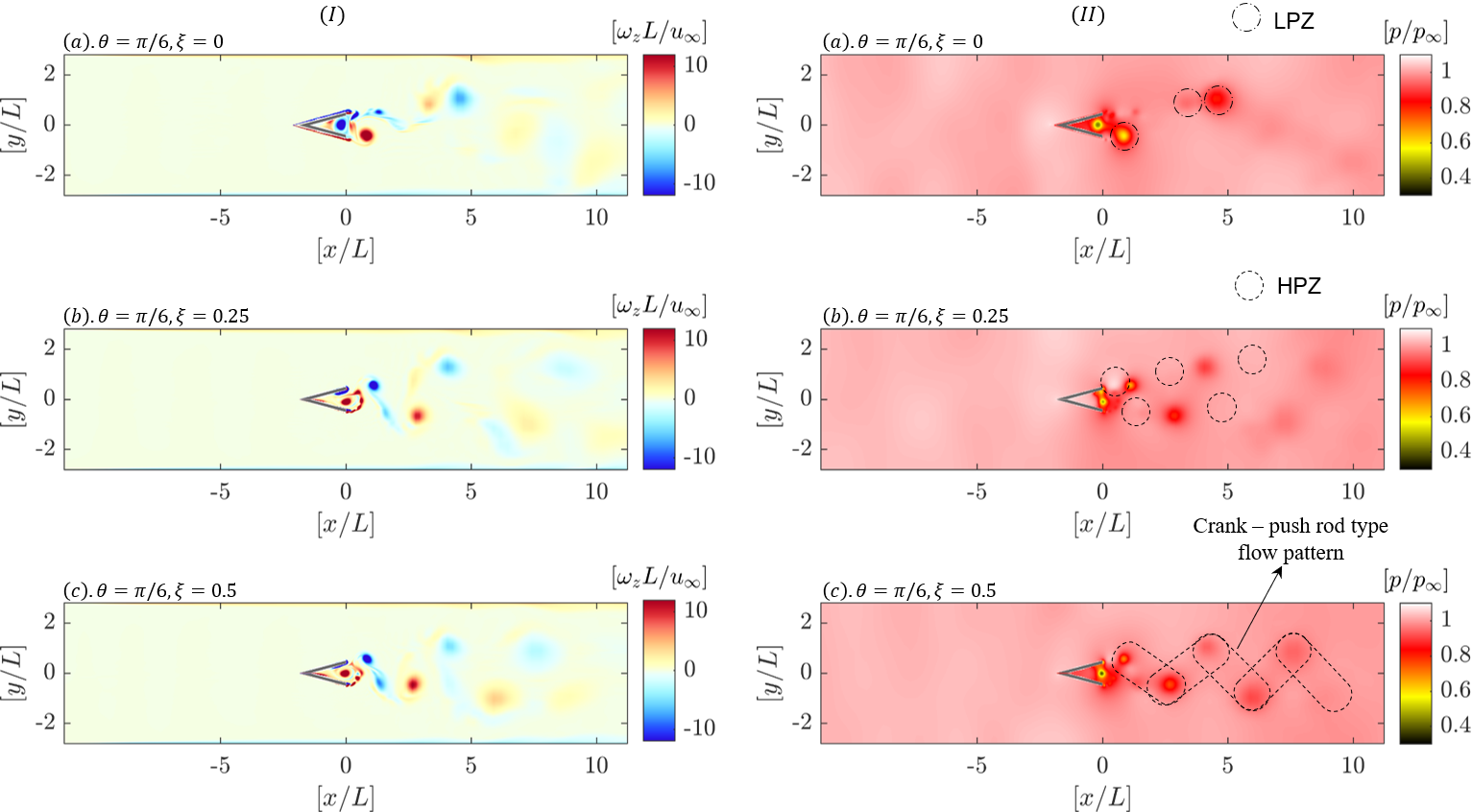}}
	\caption{\href{https://youtu.be/iyVkpbSkG1o}{(Multimedia view)} Instantaneous contour plots observed at an arbitrary time step during the simulation of the v-gutter geometry for a particular angle in degree: $\theta=[\pi/6]$ with different slit sizes in mm: $\xi=[0, 0.25,0.5]$. (I) Non-dimensionlized vorticity about the z-axis $[\omega_{z}L/u_\infty]$ and (II) Non-dimensionlized  static pressure $[p/p_\infty]$. Vorticity is non-dimensionlized using v-gutter's width of $L=17.8$ mm and freestream velocity $u_\infty =85 m/s$. The static pressure is non-dimensionlized using freestream pressure of $p_\infty =1.01\times 10^5$, Pa.}
	\label{fig:inst_vor_press_slits}
\end{figure*}

The major results from the computations are discussed under four different categories. In the first subsection ($\S$ \ref{ssec:inst_flow}), the key features and the flow field changes that are observed from the instantaneous flow are discussed. In the next section ($\S$ \ref{ssec:time_avg}), time-averaged statistics of the key flow variables that include streamwise velocity, transverse velocity, and local pressure are presented. In the third subsection ($\S$ \ref{ssec:p_loss}), total pressure loss across the duct is discussed. In the fourth subsection ($\S$ \ref{ssec:unsteady_flow}), the unsteadiness arising from the v-gutter angle changes and the slit size variations are considered. In the fifth subsection ($\S$ \ref{ssec:momentum}), momentum deficit along the duct axis is elaborated. In the last subsection ($\S$ \ref{ssec:spatiotemporal}), some of the dominant spatial modes from the proper orthogonal decomposition (POD) and the dominant temporal modes from the dynamic mode decomposition (DMD) are presented. The prominent geometrical feature that ail the desired flow field is picked after understanding the underlying physics through the analysis mentioned above. In this particular case, a flow field with a wide fluctuating spatial field downstream of the v-gutter with a robust vortex shedding at a discrete frequency and a minimum total pressure loss is the item of interest. The variations in v-gutter's geometrical parameters, including v-gutter angle ($\theta,^\circ$) and slit size ($\xi,$ mm), will be the primary focus in the upcoming sections. Hence, the dimensional units of $\theta$ and $\xi$ are omitted for discussion brevity unless specified.  
 
\subsection{Instantaneous flow field} \label{ssec:inst_flow}

The instantaneous flow field around the v-gutter is explained through the vorticity ($\omega_z$) and pressure ($p$) contours. The flow variables are selected to represent the combined effects of velocity and pressure. The flow vorticity (see Figure \ref{fig:inst_vor_press_angle}-I) is normalized by the minimum eddy turn around the time-scale given by the freestream velocity ($u_\infty$) and the transverse width ($L$) of the v-gutter. The pressure field (see Figure \ref{fig:inst_vor_press_angle}-II) is normalized by the freestream static pressure ($p$). Firstly, the effects of the v-gutter angle ($\theta$) are probed. Qualitatively, as $\theta$ increases, the vortex shedding in the wake moves from aperiodic to periodic. The behavior of aperiodicity is prevalent only for $\theta=[\pi/6]$ and mild for $\theta=[\pi/4]$. As $\theta$ increases further, the vortex shedding becomes periodic. Furthermore, the wavelength of the shedding vortex almost remains the same with $\lambda/L\sim 3.3$. 

The periodic behavior can be visually appreciated much in the pressure contour (Figure \ref{fig:inst_vor_press_angle}-II). The presence of a crank-push rod type flow pattern (von Karman vortex street) between $[\pi/3]\leq \theta \leq [\pi/2]$ ensures the presence of a periodic vortex shedding. Such kind of flow pattern is absent or weakly visible in $[\pi/6]\leq \theta \leq [\pi/4]$. Moreover, the upstream part of the duct is dominated by pressure fluctuations. The upstream fluctuations are seen to be increasing gradually as $\theta$ increases. The flow curls up strongly at either side of the v-gutter's shoulder tip. Two opposite vortices are thus, formed in the wake in an alternative manner. The flow in between the vortices accelerates, leading to the presence of low-pressure zones (LPZ). However, as the flow exits the vortex canal, it interacts with the incoming accelerated flow from the shoulder and stagnates. The stagnating fluid experiences a local pressure rise seen as a bright white pocket of a high-pressure zone (HPZ). The HPZ bursts open and relieves the surrounding fluid's pressure as it convects. The resulting micro-pressure pulse is felt everywhere in the duct. It is prominently seen in the upstream part of the duct owing to the absence of significant flow features as seen in the wake. 

In Figure \ref{fig:inst_vor_press_slits} I-II, the vorticity and pressure contours for different slit sizes ($\xi$) are given. Qualitatively, from the vorticity contours (Figure \ref{fig:inst_vor_press_slits}-I), for the plain case ($\xi=0$), aperiodicity is prevalent as discussed before. However, as the slit size increases, the intensity of aperiodicity decreases, and the vortices in the wake are seen to be shedding in an almost orderly manner. The weak presence of crank-push rod flow pattern in the pressure contours (Figure \ref{fig:inst_vor_press_slits}-II) of $\xi=0.5$ case explains the weakening aperiodicity strength and vice versa for the other cases of $\xi$. Similarly, the influence of $\xi$ increases the overall fluctuations seen in the duct in contrast to the effect of $\theta$. A thorough analysis is further needed to quantify the intensities systematically, which is done in the upcoming sections.

\begin{figure*}
	\centering{\includegraphics[width=0.9\textwidth]{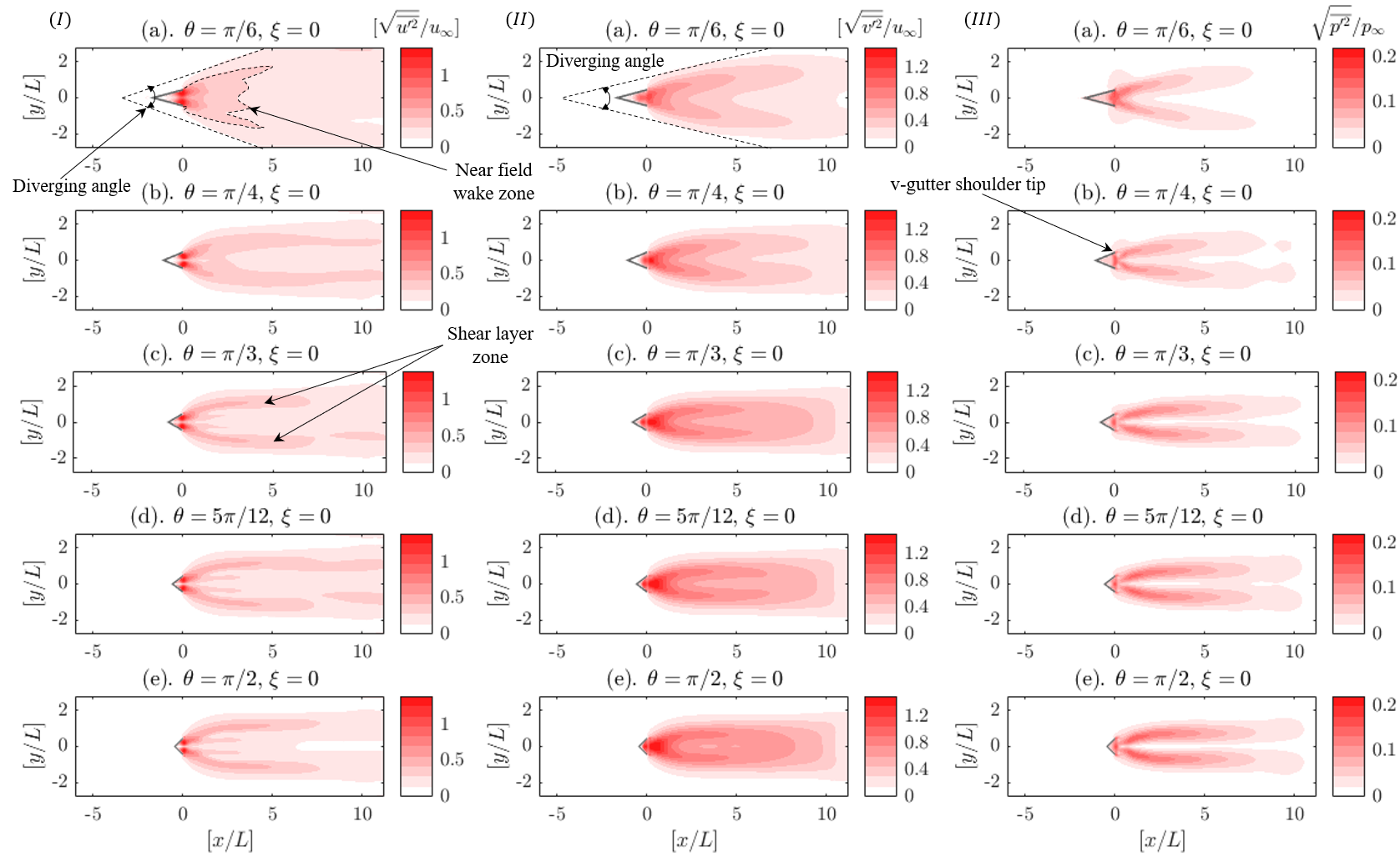}}
	\caption{Typical dimensionless fluctuation intensity contour plots observed for different v-gutter angles in degree: $\theta=[\pi/6,\pi/4,\pi/3,5\pi/12,\pi/2]$ at $\xi=0$. The fluctuation intensities are considered for three different flow parameters: (I) streamwise velocity fluctuation intensity $[\sqrt{\overline{u'^2}}/u_\infty]$ (II) transverse velocity fluctuation intensity $[\sqrt{\overline{v'^2}}/u_\infty]$ (III) static pressure fluctuation intensity $[\sqrt{\overline{p'^2}}/p_\infty]$.}
	\label{fig:fluc_intensities_angle}
\end{figure*}

\subsection{Time-averaged flow field} \label{ssec:time_avg}

\begin{figure*}
	\centering{\includegraphics[width=0.9\textwidth]{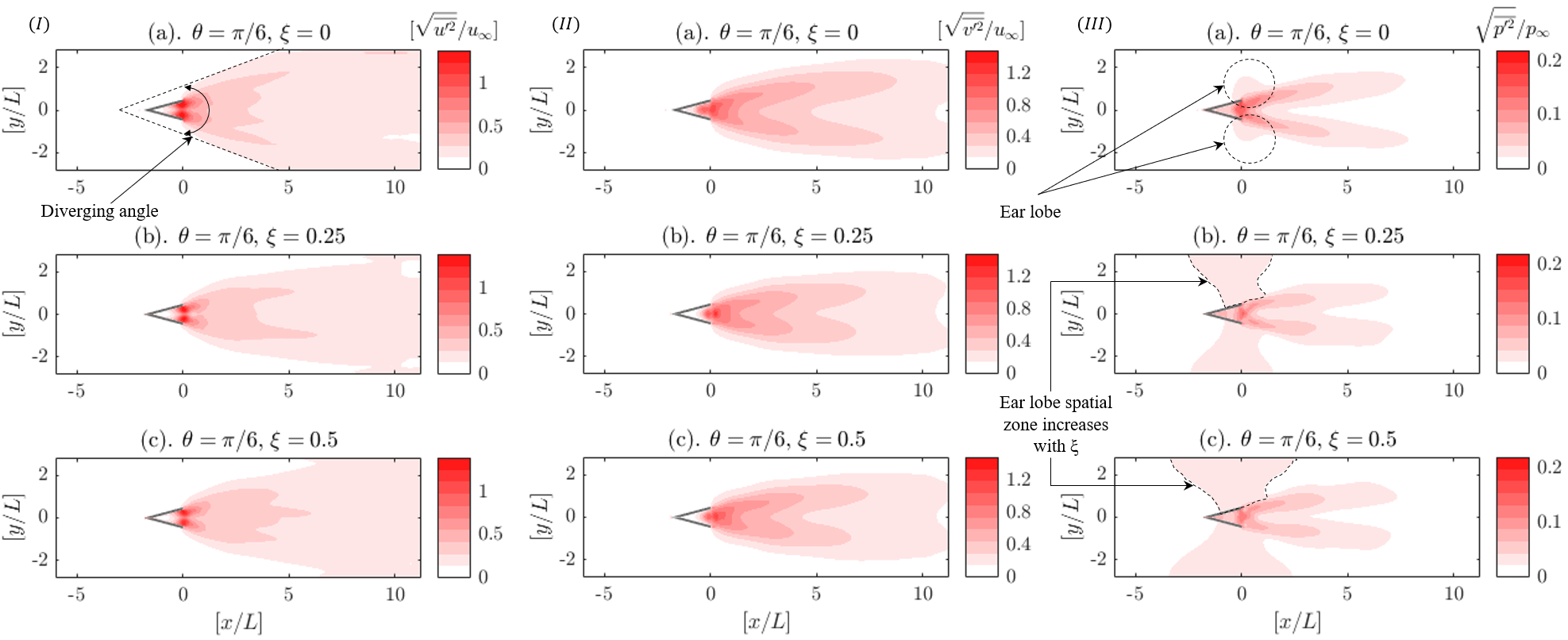}}
	\caption{Typical dimensionless fluctuation intensity contour plots observed for a v-gutter with different slit sizes in mm: $\xi=[0, 0.25,0.5]$ at $\theta=[\pi/6]$. The fluctuations are considered for three different flow parameters: (I) streamwise velocity fluctuation intensity $[\sqrt{\overline{u'^2}}/u_\infty]$ (II) transverse velocity fluctuation intensity $[\sqrt{\overline{v'^2}}/u_\infty]$ (III) static pressure fluctuation intensity $[\sqrt{\overline{p'^2}}/p_\infty]$.}
	\label{fig:fluc_intensities_slit}
\end{figure*}

The gross features of the wake flow and fluctuations in the duct are assessed using the fluctuation intensities of streamwise velocity ($u$), transverse velocity ($v$), and pressure ($p$). The fluctuation in velocity and pressure field help understand the turbulent mixing progression in the wake and the overall acoustics of the duct, respectively. In addition, the former helps identify the zones of air-fuel mixing in practical scenarios, whereas the latter aids in controlling the flame-related instabilities during combustion. As a general comment, it has to be said that in most cases, the features of spatial fluctuations under consideration are not exactly symmetric. One reason is the need to have an extended flow time to compute perfect higher-order statistics. As it is not the case in the present discussion due to computational power limitations, only broad spatial features of the higher-order statistics are evaluated.

In Figure \ref{fig:fluc_intensities_angle}-I, the normalized streamwise fluctuation intensity for each cases of $\theta$ is given. For $\theta=[\pi/3]$, due to aperiodicity, the near wakefield is dominated by strong fluctuations. However, as $\theta$ increases ($[\pi/4] \leq \theta \leq [\pi/2]$), fluctuations in the near wake zones are reducing. Moreover, the fluctuations along the shear layer arising in the v-gutter's shoulders are prominent. For $\theta > [\pi/3]$, irrespective of $\theta$, the spread of the fluctuations along the shear layer is contained only to a total width of about $||[\delta (y)/L]||\sim 4$ at $[x/L] \sim 11$. In Figure \ref{fig:fluc_intensities_slit}-I, the effect of slit ($\xi$) is shown. As $\xi$ increases, instantly, the diverging angle of the fluctuating spatial field starts to decrease between $\xi = 0$ and 0.25. Whereas, for $\xi = 0.25$ and 0.5, the diverging angle increases slightly. The near wake zone also expands gradually. The fluctuations along with the shear layer are eventually diminishing and become indistinguishable in the wake zone, especially for $\xi=0.5$. 

In Figure \ref{fig:fluc_intensities_angle}-II, the normalized transverse fluctuations are given. As $\theta$ increases, the fluctuations contained along the shear layer are distributed throughout the wake. For $\theta=[\pi/2]$, the entire wake is dominated by transverse fluctuations. However, while analyzing the influence of slit (Figure \ref{fig:fluc_intensities_slit}-II), the transverse fluctuations are seen to be damped and contained slightly closer to the v-gutter as $\xi$ increases. Although contained, even for $\xi=0.25$ and 0.5, the fluctuations are seen predominantly in the shear layer.

In Figure \ref{fig:fluc_intensities_angle}-III, the pressure fluctuations are shown. The flow velocity accelerates at the v-gutter's shoulder tip, and the exit angle of the slipping velocity field increases with $\theta$. The resulting horizontal velocity component thus decreases with increasing $\theta$. For shallow $\theta$, the pressure fluctuation is higher around the shoulder as the horizontal velocity is observed to be the highest among the considered cases. A small lobe-like feature present around the shoulder explains the extent of pressure fluctuation for $\theta=[\pi/6]$.

In the cases of slits (Figure \ref{fig:fluc_intensities_slit}-III), as $\xi$ increases, the ear lobe spatial zone increases both upstream and downstream of the v-gutter's shoulder progressively. The damping of aperiodic vortex shedding at $\xi=0$ and almost periodic vortex shedding at $\xi = 0.25$ and 0.5 are attributed to earlier behavior. However, the aperiodic shedding is only switched to a nearly periodic condition due to the mass and momentum leakage in the form of a jet through the leading edge slit. The diffused jet flows through the slit and further flaps inside the v-gutter. While flapping, the jet coming out of the slit attaches to the inside edge of the shoulder opposite the vortex ejecting edge of the considered cycle. Such flapping motion periodically forces the shoulder tip with a transverse velocity. The resulting periodic forcing amplifies the pressure fluctuation and thereby leading to the observation of more prominent earlobes for $\xi = 0.25$ and 0.5.

\subsection{Total pressure loss across the duct} \label{ssec:p_loss}
\begin{figure*}
	\centering{\includegraphics[width=0.9\textwidth]{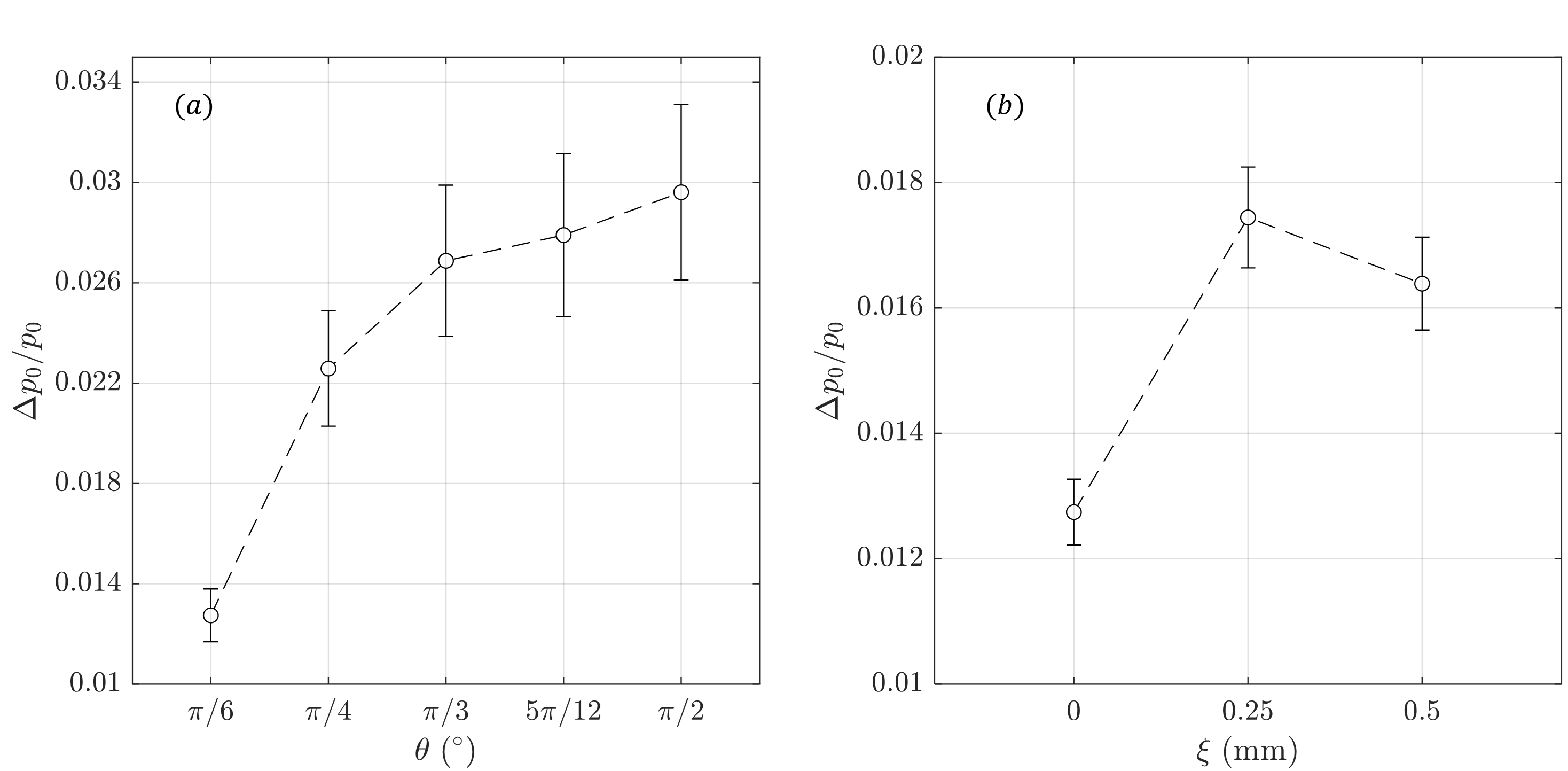}}
	\caption{Graph showing the total pressure loss ($\Delta p_0/p_0$) between the entry and exit of the duct with different geometrical types of v-gutter: (a) variations in v-gutter's angle in degree: $\theta=[\pi/6,\pi/4,\pi/3,5\pi/12,\pi/2]$) and (b) variations in v-gutter's slit size in mm for $\theta=[\pi/6]$: $\xi=[0, 0.25,0.5]$. The error bar denotes the scaled and non-dimensionalized total pressure fluctuation intensity encountered at the exit ($0.1\sigma_{p_0}/p_0$).}
	\label{fig:p_loss}
\end{figure*}

One of the primary concerns for a compressible flow in a duct is the total pressure loss ($\Delta p_0$). A typical variation of the non-dimensionalized $\Delta p_0$ is shown in Figure \ref{fig:p_loss} for different $\theta$ and $\xi$ variations. In the present problem, $\Delta p_0$ is calculated by computing the line average of $p_0$ at the entry and exit of the duct. Later, the difference ($\Delta p_0$) is normalized to $p_0$ to quantify the total pressure loss as a fraction. The error bar is used as an indicator to tell the extent of $p_0$ fluctuations ($\sigma_{p_0}$) in the exit owing to the presence of the v-gutter. For readability in Figure \ref{fig:p_loss}, $\sigma_{p_0}$ is rescaled appropriately. From Figure \ref{fig:p_loss}a, losses due to the v-gutter's $\theta$ can be understood. The flow is chaotic in the wake of the aperiodic case ($\theta = \pi/6$). However, due to the streamlined body shape of the v-gutter with low $\theta$, losses in $p_0$ are almost minimal (~1.25\%) compared to the other cases. As $\theta$ increases, $\Delta p_0$ increases rapidly. Even for the almost periodic case ($\theta=\pi/4$), ~75\% increment in $\Delta p_0$ is seen, which is quite high. As $\theta$ increases ($\theta>\pi/4$) further, wake structures become more periodic and variations in the $\Delta p_0$ are minimal (~15\%). However, in comparison to the streamlined case ($\theta=\pi/6$), the most bluff body ($\theta=\pi/2$) case encounters a substantial increment in $\Delta p_0$ (~140\%). 

The presence of a slit minimizes the high-pressure loss in the wake due to the penetration of mass through it. From Figure \ref{fig:p_loss}b, as $\xi$ changes, pressure loss increases quickly for $\xi=0.25$ by ~26\% and drops slightly for $\xi=0.5$ by ~5\% in comparison with $\xi=0.25$. For a narrow slit case ($\xi=0.25$), the viscous forces dominate and contribute to the higher $\Delta p_0$. Whereas for the wide slit case ($\xi=0.5$), the loss due to viscous forces is comparatively less. However, in the presence of a slit ($\xi=[0.25,0.5]$), the fluctuation in $p_0$ or the standard deviation of the $p_0$ profile in the duct's exit $(\sigma_{p_0(y)})$ is comparatively higher than the plain case. It remains almost invariant with variations in slit size ($\xi$). Thus, in terms of flow mixing performance, one can appreciate the presence of lower $\Delta p_0$ and an almost periodic flow for a shallow v-gutter angle with a wide slit.  

\subsection{Unsteady dynamics} \label{ssec:unsteady_flow}

\begin{figure*}
	\centering{\includegraphics[width=0.9\textwidth]{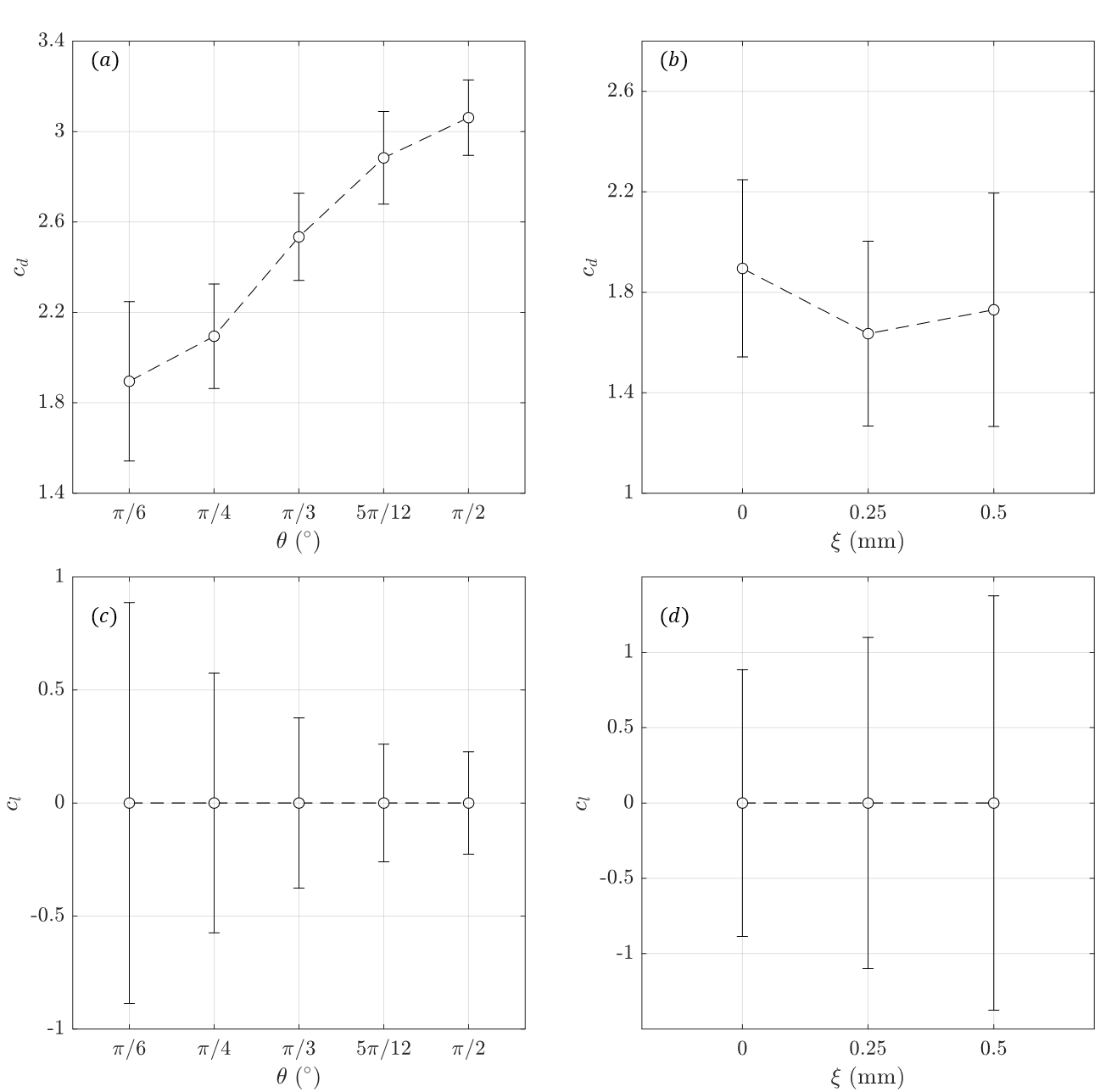}}
	\caption{Graph showing the variation in overall drag $(c_d)$ and lift $(c_l)$ coefficients for different v-gutter configurations: (a,c) effects of v-gutter's angle in degree- $\theta=[\pi/6,\pi/4,\pi/3,5\pi/12,\pi/2]$). (b,d)  effects of v-gutter's slit sizes in mm-$\xi=[0, 0.25,0.5]$ at $\theta=[\pi/6]$. The errorbar indicates the fluctuation intensity ($\sigma$) in the $c_l$ and $c_d$ values and are represented as rescaled values of $0.5\sigma_{c_l}$ and $0.5\sigma_{c_d}$ for clarity.}
	\label{fig:cl_cd_variations}
\end{figure*}

\subsubsection{influence on the overall forces}
The effects of $\theta$ and $\xi$ play a vital role in the overall forces acting on the v-gutter. Being a two-dimensional model, the non-dimensional force coefficients underplay are drag ($c_d$) and lift ($c_l$). V-gutters contribute significantly to the total drag acting on the duct flow as they are the bluff bodies aside from the skin friction drag on the wall. In Figure \ref{fig:cl_cd_variations} a-b, the variation of $c_d$ in terms of $\theta$ and $\xi$ are given. In general, the value of $c_d$ increases with $\theta$ almost linearly. On the other hand, as $\xi$ changes $c_d$ drops quickly for $\xi=0.25$ and increase slightly for $\xi=0.5$. However, the fluctuations to the mean remain the same for $\theta$ and $\xi$ variations, except for $\theta=[\pi/6]$. As $\theta$ increases from $[\pi/6]$ to $[\pi/2]$, a total increment of 63\% in $c_d$ is observed. In terms of the fluctuations, between $\theta=[\pi/6]$ and $[\pi/4]$, a total drop of 42\% is seen, and it remains almost constant for all the cases of $\theta$. A maximum drop-in $c_d$ of about 16\% is observed between $\xi=0$ and 0.25. However, a drop of about 11\% is seen between $\xi=0$ and 0.5. The higher drag for higher $\theta$ is explained by the diverging angle observed at the v-gutter's shoulder. A high diverging angle due to high $\theta$ separates the flow efficiently as the streamlines leave the shoulder to offer a large wake zone. Pressure drag tends to increase when the wake zone is prominent. As the presence of a slit immediately dissipates the high-pressure formation in the large wake zone, the overall pressure drag decreases considerably. The slight increment in $c_d$ between $\xi=0.25$ and 0.5 cannot be explained with the current analysis. However, it is suspected due to the way the jet diffuses through the slit based on the slit size or $\xi$.

In Figure \ref{fig:cl_cd_variations} c-d, the influence of $c_l$ is discussed with respect to $\theta$ and $\xi$. As the v-gutter is at zero angle of attack, there is no net $c_l$ observed on the model. However, the level of fluctuations seen during the considered cycles varies for different $\theta$ and $\xi$. For the aperiodic case of $\theta=[\pi/6]$, the fluctuations are observed to be a maximum of 90\%. As $\theta$ increases, the fluctuation intensity gradually decreases to the least value of 25\%. On the contrary to $\theta$ influence, as $\xi$ increases, a progressive increment in the fluctuation intensity of $c_l$ is seen. A maximum of 56\% increment is seen between the cases of $\xi=0$ and 0.5. Firstly, the decrements in $c_l$ for $\theta$ are explained by lesser v-gutter length for high $\theta$. The lifting surface length can be approximated as the distance upstream from the origin to the point where the tip extends. As the lift-producing surface decreases with an increase in $\theta$, the overall fluctuation intensity decreases with $\theta$. On the other hand, the diffusivity and flapping intensity of the jet from the slit increase with $\xi$. Although the lifting surface lengths are constant for different cases of $\xi$, the flapping of the diffusing jet increases the fluctuation intensity of $c_l$ as $\xi$ increases. 

\subsubsection{Flow fluctuations in the duct} \label{ssec:global_fluc}

\begin{figure*}
	\centering{\includegraphics[width=0.95\textwidth]{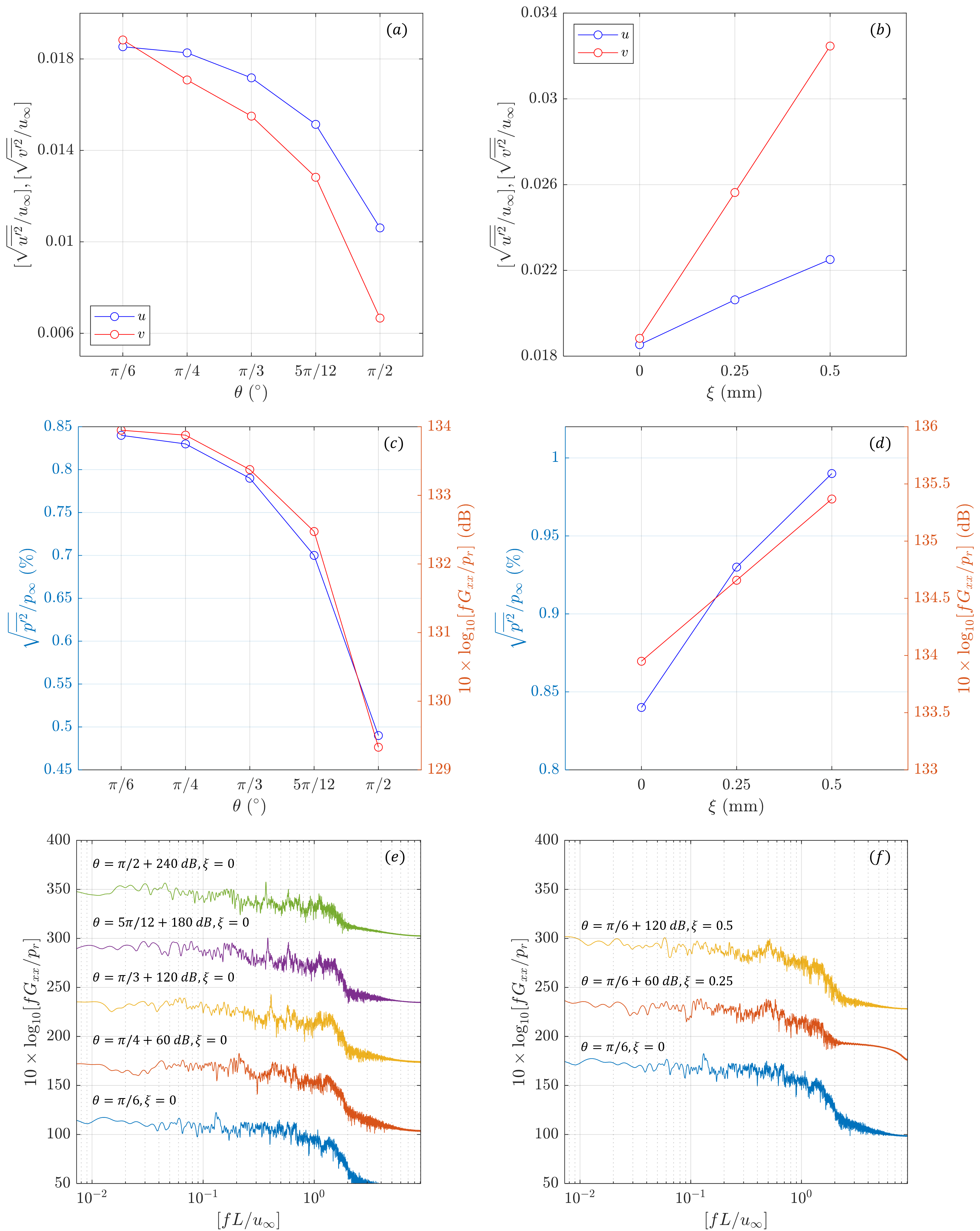}}
	\caption{Graph showing the upstream fluctuation intensities for different flow parameters including streamwise velocity $(u)$, transverse velocity $(v)$, static pressure $(p)$ and pressure fluctuation spectra at $[x/L]=-5$ and $[y/L]=0$. Maximum fluctuation intensities and pressure spectra for different v-gutter geometries: (a,c,e) for different angles in degree, $\theta=[\pi/6,\pi/4,\pi/3,5\pi/12,\pi/2]$) at $\xi=0$ mm, (b,d,f) for different slit sizes in mm: $\xi=[0, 0.25,0.5]$ at $\theta=[\pi/6]$. To avoid cluttering of spectra in (e,f), certain offset is provided for each of the cases.}
	\label{fig:upstream_fluc_plot}
\end{figure*}

In Figure \ref{fig:upstream_fluc_plot}, the influence of v-gutter's $\theta$ and $\xi$ in terms of flow fluctuations are plotted. Firstly, through Figure \ref{fig:upstream_fluc_plot} a-b, the influence of velocity ($u,v$) fluctuations are discussed. Secondly, using Figure \ref{fig:upstream_fluc_plot} c-d, outcomes from the representative pressure ($p$) fluctuations and overall sound pressure level (OSPL, dB) are discussed. At last in Figure \ref{fig:upstream_fluc_plot} e-f, upstream pressure fluctuation spectra are discussed. In all the results, a point at $[x/L,y/L]=[-5,0]$ is taken as an indicative point representing the global fluctuations seen in the upstream side of the duct. The fluctuation intensity gives a broad idea of the extent of combustion instabilities, flame holding unsteadiness, or duct noise seen in a ramjet combustor. In general, the streamwise and transverse velocity fluctuations are observed to be decreasing as $\theta$ increases. The values decrease gradually up to $\theta=[5\pi/12]$ but drop drastically at $\theta=[\pi/2]$. In summary, a maximum decrements of about 44\% and 66\% in both streamwise and transverse fluctuations as $\theta$ increases from $[\pi/6]$ to $[\pi/2]$, respectively. On the contrary, as $\xi$ increases, the values of streamwise and transverse velocity fluctuations increase almost linearly. A total increment of about 22\% and 83\% in the streamwise and transverse velocity fluctuations is seen as $\xi$ increases.

Aside from quantifying velocity fluctuations (which are suitable for understanding the general instabilities observed in the duct in terms of flow kinematics), monitoring the changes in the local pressure field is of paramount importance in quantifying the global duct noise. Pressure fluctuations are observed at the same specific spatial point as before to generalize the generic aeroacoustics of the duct that is present upstream of the v-gutter (see Figure \ref{fig:upstream_fluc_plot} c-d). In general, the normalized pressure fluctuations to the atmospheric pressure are seen to be following the trend of velocity fluctuations. A maximum drop in the non-dimensionalized pressure fluctuation and OSPL (dB) is around 47\% and 4\%, respectively, for $\theta$ variations. Similarly, the trend for $\xi$ variations is similar to that of the velocity fluctuations. A maximum gain in the non-dimensionalized pressure fluctuation and OSPL (dB) is observed to be around 18\% and 1\%, respectively. In summary, it can be seen that the noise level is quite high upstream for the streamlined body ($\theta=\pi/6$), whereas if a slit is added, noise levels are slightly increasing. 

However, analysing the spectra from Figure \ref{fig:upstream_fluc_plot} e-f, discrete pressure loading is observed as $\theta$ increases, especially after $\theta \geq [\pi/3]$. During the initial $\theta$, the dominant peak is observed to be on the lower end ($fL/u_\infty=0.13$ for $\theta=\pi/6$). Even for $\theta=[\pi/4]$, the spectra is still on the lower side, however, compared to $\theta=[\pi/6]$, the spectra is slightly higher ($fL/u_\infty=0.4$) and broad. For $\theta=[\pi/3,5\pi/12,\pi/2]$, a discrete peak with a gradual decaying spectra is seen ($fL/u_\infty \sim 0.4,0.38,0.36$). Switching spectral signs from broad to discrete are also seen in the $\xi$ variations cases. As $\xi$ increases ($\xi=0,0.25,0.5$), the broadened spectra starts to transform into a discrete peak however with a wide bandwidth ($fL/u_\infty \sim 0.13,0.5,0.52$). The peak spectra for the cases with slits remain almost the same on the higher end. Thus, in summary, it can be said that the streamlined body ($\theta=\pi/6$) produces a low-frequency signature, and if a slit is added, the noise level is observed to be increasing slightly, and discrete spectra are seen on the higher end. Later in the upcoming sections, it will be shown that these frequencies are indeed harmonics of the dominant shedding frequency in the wake.

\begin{figure*}
	\centering{\includegraphics[width=0.9\textwidth]{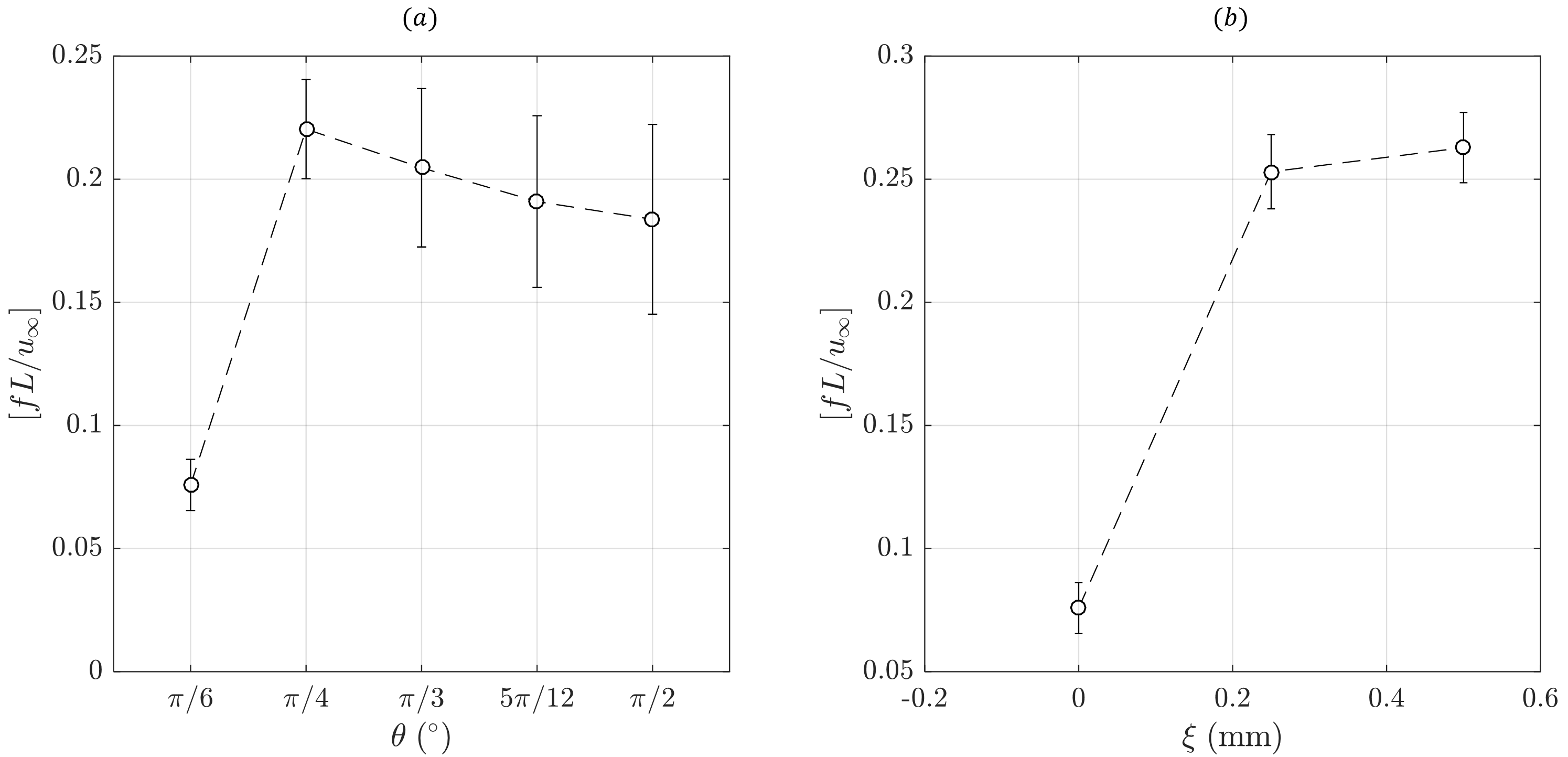}}
	\caption{Graph showing the non-dimensional frequency ($fL/u_\infty$) variation observed in the pressure contours for different v-gutter configurations at $[x/L]=5$ and $[y/L]=-0.5$. For non-dimensionalization, v-gutter's shoulder width is taken as the reference length scale ($L=17.8$ mm), and the freestream velocity as the reference velocity scale ($u_\infty=85$ m/s). Non-dimensionalized frequency variation is represented for different v-gutter geometries: (a) for different angles in degree, $\theta=[\pi/6,\pi/4,\pi/3,5\pi/12,\pi/2]$) at $\xi=0$ mm, (b) for different v-gutter's slit sizes in mm: $\xi=[0, 0.25,0.5]$ at $\theta=[\pi/6]$. The errorbar represents the overall pressure fluctuation intensity ($\sqrt{\overline{p'^2}}/p_\infty$).}
	\label{fig:spectra_angle_slits}
\end{figure*}

The previous section has already explained that the effect of $\theta$ increment reflects in reducing the streamwise velocity component and increasing the transverse velocity components near the v-gutter's shoulder tip. The resulting kinematic field play thus alters the way the vortices are convected downstream, notably at a lower shedding rate. The lower the vortices are shed with a lower streamwise component in the shoulder tip, the lesser the pressure buildup between the vortices and thus the lesser the fluctuations when they burst. Detailed analysis of the vortex shedding frequency needs to be done to verify whether such an event exists.

\subsubsection{Vortex shedding frequency} \label{ssec:vortec_shed}

The global vortex shedding frequency behind the v-gutter's wake is assessed by monitoring the pressure fluctuation. Along the streamwise direction, where the interface of the considered near and far-field zone exists (see Figure \ref{fig:inst_vor_press_angle}c), a point is taken along the horizontal line passing through the v-gutter's shoulder tip: $[x/L,y/L]=[5,-0.5]$. The point is chosen as it represents the pressure fluctuations seen during the vortex shedding period for all the cases under consideration. The variation of the dominant frequency seen in each of the cases under investigation is plotted in Figure \ref{fig:spectra_angle_slits}. Firstly, the effects of $\theta$ are explained in Figure \ref{fig:spectra_angle_slits}a. For $\theta=[\pi/6]$, the dominant non-dimensional frequency is observed to be the lowest at $[fL/u_\infty]\sim0.076$. The details of other spectra observed in this case will be explained in the upcoming section, where a discussion is given using the $x-t$ diagram. 

Asides, as $\theta$ increases, $[fL/u_\infty]$ decreases gradually in almost a linear manner. A total decrement of about $\sim17\%$ is encountered as $\theta$ changes from $[\pi/4]$ to $[\pi/2]$. The sudden jump in $[fL/u_\infty]$ for initial increments in $\theta$ and a subsequent decrement in $[fL/u_\infty]$ for higher $\theta$ is due to two reasons: one is due to the fact the shedding becomes more organized and the second is because of the decrement in the streamwise convective velocity away from the v-gutter's shoulder tip for higher $\theta$. Contrary to the previous, $[fL/u_\infty]$ values tend to increase as $\xi$ increases. However, there is no definitive trend seen. A total increment in $[fL/u_\infty]$ is seen to be $\sim65\%$ as $\xi$ increases from 0 to 0.5. The diffusion of fluid mass through the slit and the further complicated flow kinematics that happen in the v-gutter's wake ensure a somewhat organized vortex shedding compared to the other cases.

In Figure \ref{fig:spectra_angle_slits}, an errorbar is marked at each frequency. The length of the error bar represents the overall pressure fluctuation intensity ($\sqrt{\overline{p'^2}}/p_\infty$) observed at the measurement location. The fluctuation intensity provides information on the extent of variations and the presence of dominant frequency. For example, in the case of $\theta$ variation, the fluctuation intensity is low for the aperiodic case of $\theta=[\pi/6]$. However, the value is at the maximum for the most periodic case $\theta=[\pi/2]$. Unlike the non-dimensional frequency $[fL/u_\infty]$, fluctuation intensity increases monotonically. A maximum increment of 2.7 times in fluctuation intensity is seen between the cases of $\theta = [\pi/6]$ and $[\pi/2]$. As $\xi$ increases, the fluctuation intensity values are also increasing (between $\xi=0$ and 0.25) and remain almost constant ($\xi=0.5$). The periodic shedding behavior might be one of the reasons. However, the increment is not as pronounced as in the $\theta$ variations. A total increment of $\sim45$\% is seen in the fluctuation intensity values between $\xi=0$ and 0.25.

\subsubsection{Wake vortices convection and spectral analysis} \label{xt_xf_analysis}

\begin{figure*}
	\centering{\includegraphics[width=1\textwidth]{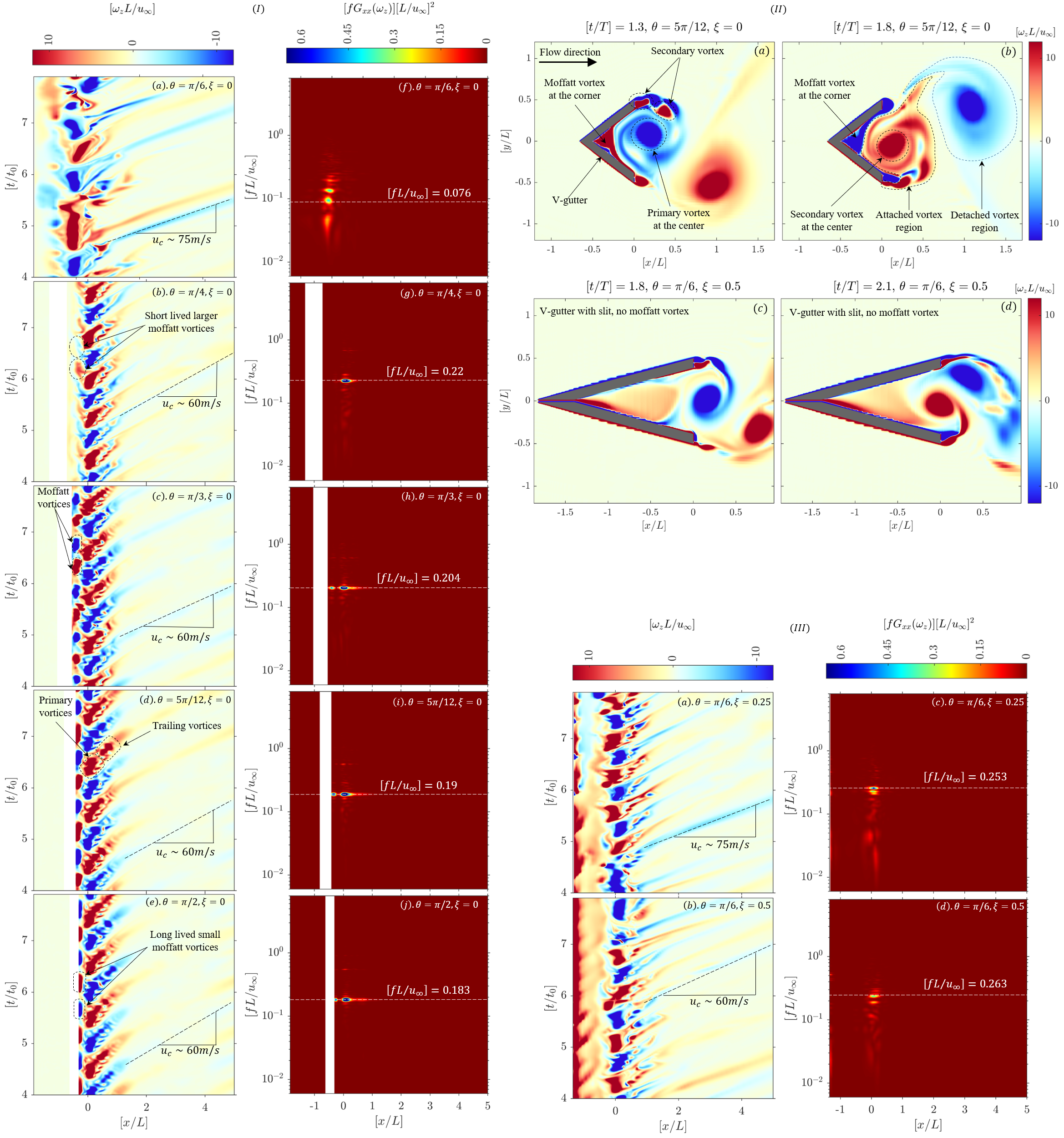}}
	\caption{(I) \href{https://youtu.be/lVrROwjga7E}{(Multimedia View)} Graph showing the contour plots of non-dimensionlized vorticity about the z-axis $[\omega_{z}L/u_\infty]$ and non dimensionlized power spectral density $[fG_{xx}(\omega_{z})][L/u_\infty]^2$ along the centerline ($y/L=0$) for v-gutter's different geometrical configurations. Varying v-gutter's angles in degree: (a,f). $\theta=\pi/6,\xi=0$ mm, (b,g). $\theta=\pi/4,\xi=0$ mm, (c,h). $\theta=\pi/3,\xi=0$ mm, (d,i). $\theta=5\pi/12,\xi=0$ mm, (e,j). $\theta=\pi/2,\xi=0$ mm. (II) A typical vorticity contour plot observed at two different extrema of a cycle for different v-gutter configurations: (a,b). $\theta=5\pi/12,\xi=0$ mm and (c,d). $\theta=\pi/6,\xi=0.5$ mm. (III) \href{https://youtu.be/oz7VugCnp4g}{(Multimedia View)} Graph showing the contour plots of non-dimensionlized vorticity about the z-axis $[\omega_{z}L/u_\infty]$ and non-dimensionlized power spectral density $[fG_{xx}(\omega_{z})][L/u_\infty]^2$ along the centerline ($y/L=0$) of the v-gutters with different slits sizes: (a,c). $\theta=\pi/6,\xi=0.25$ mm, and (b,d). $\theta=\pi/6,\xi=0.5$ mm.}
	\label{fig:xt_pic}
\end{figure*}

The variations of $\theta$ and $\xi$ on the wake-vortex-dynamics are studied in depth using the $x-t$ and $x-f$ diagrams. The $x-t$ diagram provides information on the temporal evolution of a particular flow structure and its convection velocity ($u_c$). The plot is between a spatial feature varying in time, and hence, a slope inverse of any streak seen in the plot gives us $u_c$ directly. In the present analysis, a line segment passing through $[y/L]=0$ is taken, and the variations of lateral vorticity ($\omega_z$) are constructed to time. Later, the constructed $x-t$ diagram can be used to perform a fast-Fourier transform (FFT) analysis to extract the dominant spectral contents that vary with $[x/L]$ ($x-f$ diagram). Typical $x-t$ and $x-f$ diagrams for different $\theta$ and $\xi$ can be seen in Figure \ref{fig:xt_pic} I and III. A typical vorticity contour plot that shows a series of evolving near-field vortical structures at two different extremes of a typical cycle is shown for reference in Figure \ref{fig:xt_pic}-II for two of the cases.

The contour plots in Figure \ref{fig:xt_pic}-II (a-b) give information on the different wake vortices seen in the v-gutter flow problem. From the shoulder tip, an incipient or secondary vortex originates. As time evolves, the vortex grows and convects as the primary vortex. As the vortex grows along the transverse direction between the shoulder tip, the disturbance also induces `Moffatt' vortex\cite{Moffatt_1964,Shankar_1998,Polychronopoulos_2018} in between the v-gutter's corner. The Moffatt vortex rotates opposite that of the disturbance from the primary vortex. Secondary and tertiary Moffatt vortices keep forming at opposite signs approaching the corner. However, the sizes are decreasing exponentially and dissipate at the viscous scale. In the present simulation, only the primary Moffatt vortex is captured. The major reason for the occurrence of Moffatt vortices is the dominance of viscous terms in the wake, which induce such a characteristic recirculation in the opposite direction. As the v-gutter's $\theta$ decreases, the typical Moffatt vortex lifetime increases in a typical cycle. At $\theta = [\pi/6]$, the absence of the Moffatt vortex can also be correlated to the observed aperiodic wake vortex shedding as they are interdependent. At $\theta = [\pi/4]$, a short-lived Moffatt vortex is seen, and the flow looks organized compared to the previous case. The statements mentioned above are reflected appropriately in the $x-t$ plots of Figure \ref{fig:xt_pic}-I (a-e). 

The spectral contents of the $x-t$ plot of Figure \ref{fig:xt_pic}-I (a-e) are given as $x-f$ plot (non-dimensionalized) in Figure \ref{fig:xt_pic}-I (f-j). The dominant spectra in these cases are found to be similar to that of the spectra shown in the previous Figure \ref{fig:spectra_angle_slits}a. For the case of $\theta = [\pi/6]$, the spectra are broadened due to aperiodic vortex shedding. The periodic balance between the inertial and viscous terms in the wake is failed, which leads to the aperiodic shedding. Instead of the secondary vortex growing into a primary vortex by swinging in the transverse direction from the shoulder tip, the vortex gets detached and convected along the flow direction at a higher speed. The slope of the streak made by one of the convecting vortices from the $x-t$ plot indicates $u_c$ to be 76.3 m/s. The momentary lead of inertial terms in balancing the viscous terms present in the recirculating wake leads to earlier vortex detachment and rapid convection. However, as $\theta$ increases, the effective streamwise velocity at the shoulder tip decreases and tugs the inertial terms; thereby, the balance exists periodically. Once a periodic disturbance exists in the closed corner, the Moffatt vortex also originates, and the periodic shedding begins. In the periodic shedding cases ($\theta = \pi/4$ to $\pi/2$), the average convection velocity of the primary vortex is found to be $u_c \sim 60$ m/s which is about 25\% lesser than the aperiodic case ($\theta = \pi/6$).

Events of vortex shedding become interesting for the cases of $\xi$ that are shown in Figure \ref{fig:xt_pic}-III. The corresponding contour plots for the case of $\xi=0.5$ at two different time instants are shown in Figure \ref{fig:xt_pic}-II (c-d). One of the striking flow features is the absence of the Moffatt vortex in the v-gutter's corner. As the slit is open, the corner morphology of the v-gutter's geometry is lost along with the Moffatt vortex system. A highly viscous jet forms via the slit, leaking into the wake. Due to the viscous effects and the adverse pressure gradient present further downstream, the jet has a preferential attachment to the v-gutter's shoulder and flaps about the v-gutter's axis. Asymmetry introduced due to the attached jet flap does not produce a periodic shedding as seen in the non-slitted v-gutter cases ($\xi = 0$). However, the pattern in the shredded vortex is much organized. The organization is even better as the $\xi$ increases to 0.5. In fact, the respective convection velocity for $\xi = 0.5$ is of the same order ($u_c \sim 60$ m/s) as that of the $\xi=0$ cases at higher $\theta$. The shedding frequency also matches that of the values obtained in the previous Figure \ref{fig:spectra_angle_slits}b. However, the intermittency due to the presence of slit produces a broadened spectra instead of the discrete ones as seen in the periodic shedding cases. 

\subsection{Momentum thickness variation along the duct} \label{ssec:momentum}

\begin{figure*}
	\centering{\includegraphics[width=1\textwidth]{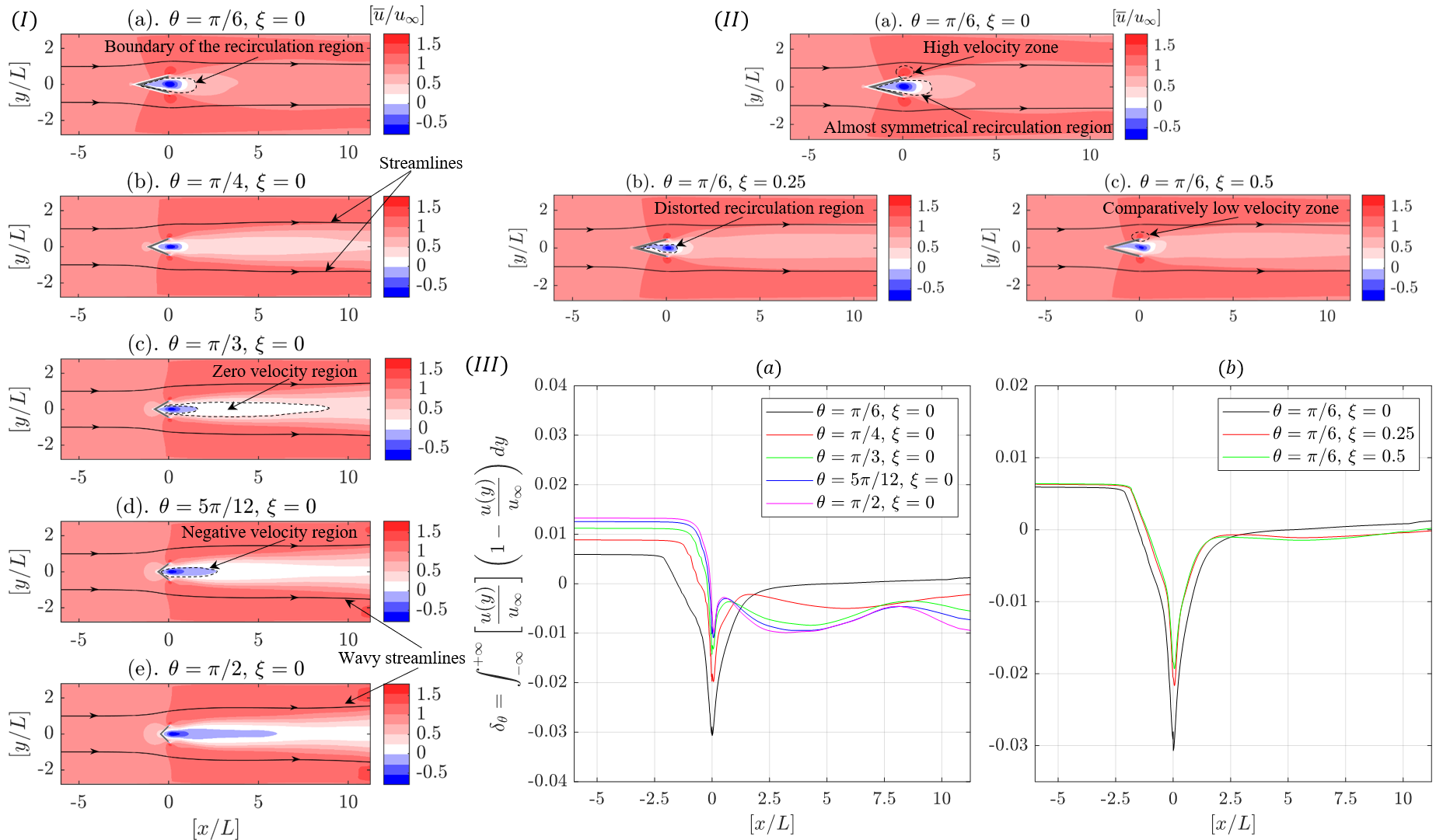}}
	\caption{Graph showing the mean streamwise velocity contour plot $[\overline{u}/u_\infty]$  with selective streamlines drawn for cross comparison between v-gutter's of different geometrical configurations: (I) variation in v-gutter's angle in degree: (a). $\theta=[\pi/6],\xi=0$ mm, (b). $\theta=[\pi/4],\xi=0$ mm, (c). $\theta=[\pi/3],\xi=0$ mm, (d). $\theta=[5\pi/12],\xi=0$ mm, and (e). $\theta=[\pi/2],\xi=0$ mm; (II) variation of v-gutter's slit size in mm: $\xi=[0, 0.25,0.5]$ at $\theta=[\pi/6]$; (III) Momentum thickness $(\delta_\theta)$ variation along the stream wise direction $(x/L)$ for v-gutter's different geometrical configurations: (a) variation in v-gutter's angle in degree-$\theta=[\pi/6,\pi/4,\pi/3,5\pi/12,\pi/2]$ at $\xi=0$ mm. (b) variation of v-gutter's slit size in mm-$\xi=[0, 0.25,0.5]$ at $\theta=[\pi/6]$.}
	\label{fig:flux_angle_slit}
\end{figure*}

The effect of confined passage and the wake dynamics of the v-gutter can be assessed though the help of the streamwise mean velocity profiles and momentum thickness as shown in Figure \ref{fig:flux_angle_slit}. Firstly, the mean velocity profile in the streamwise direction for $\theta$ variations is considered (Figure \ref{fig:flux_angle_slit}-I). In the cases of $\theta$ variation, the recirculation bubble increases monotonically with $\theta$. For the aperiodic cases ($\theta=\pi/6$, and $\pi/4$), the recirculation zone (which is characterized by the light blue contour representing the first negative velocity step) appears to be contained between $0.5\leq [x/L]\leq 1]$. As for the periodic cases (from $\theta=\pi/3$, to $\theta=\pi/2$), the recirculation region elongates rapidly between $2\leq [x/L]\leq 7]$. The streamlines drawn for each of the cases represent the flow path. The small recirculation zone does not turn the flow back to the v-gutter's axis. Whereas long recirculation zones gradually turn the flow towards the axis. The flow further interacts and recovers back to the freestream direction. Owing to this behavior, one can see a wavy streamline pattern downstream of the v-gutter for the periodic wake shedding cases. Such behavior will come in handy to explain the trends in the momentum thickness variations in the upcoming discussions. In summary, a short recirculation region is a sign of a weak wake, whereas a long recirculation zone represents the presence of an intense wake. From the previous analysis, it is clear that the cases of strong wake also exhibit persistent periodic shedding. 

Secondly, the mean velocity profile along the $x$-axis is taken for $\xi$ variations (Figure \ref{fig:flux_angle_slit}-II). In the cases of $\xi$ variations, the recirculation bubble looks collapsed, distorted and weak with increments in $\xi$. The presence of the slit tries to diminish the recirculation bubble. However, due to the unsteadiness in the wake, the bubble is only distorted and shrunk in size. Due to the bifurcation of fluid mass through the slit, the net acceleration in the tip of the shoulder also reduces as $\xi$ increases. The gross observation of the streamlines shows a minor wavy nature downstream of the v-gutter, as seen in the previous cases where periodic shedding is observed. From the vorticity contours shown in Figure \ref{fig:inst_vor_press_slits}-I b-c, the almost periodic vortex shedding can also be seen for the slitted cases ($\xi = 0.25$ and 0.5).

\begin{figure*}
	\centering{\includegraphics[width=0.9\textwidth]{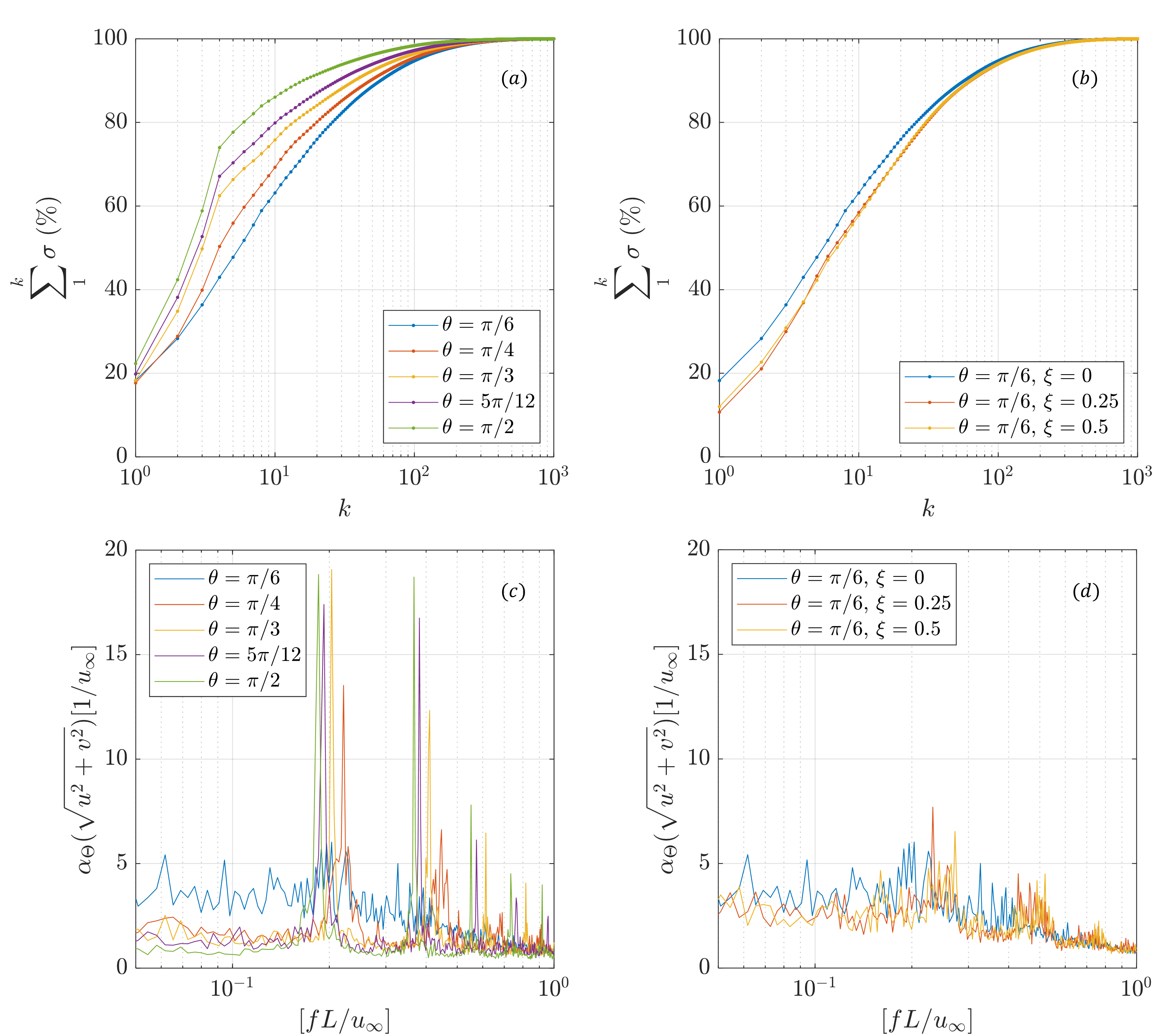}}
	\caption{Graph showing the cumulative energy distribution across the $k^{th}$ mode from the POD analysis of the velocity field for v-gutter's different geometrical configurations: (a) variation of v-gutter's angle in degree-$\theta=[\pi/6,\pi/4,\pi/3,5\pi/12,\pi/2]$ for $\xi=0$ mm; (b) variation of v-gutter's slit size in mm-$\xi=[0, 0.25,0.5]$ for $\theta=[\pi/6]$. Typical amplitude spectra from the DMD analysis of the velocity field ($fL/u_\infty vs. \alpha_\Theta(\sqrt{u^2+v^2})[1/u_\infty]$) for v-gutter's different geometrical configurations: (c) variation of v-gutter's angle in degree-$\theta=[\pi/6,\pi/4,\pi/3,5\pi/12,\pi/2]$ at $\xi=0$ mm; (d) variation of v-gutter's slit size in mm-$\xi=[0, 0.25,0.5]$ for $\theta=[\pi/6]$.}
	\label{fig:temporal_mode}
\end{figure*}

Lastly, the momentum thickness ($\delta_\theta$) along the streamwise direction is considered (Figure \ref{fig:flux_angle_slit}-III). The definition of $\delta_\theta$ is given by Eq. \ref{eqn:mom_thick} as,
\begin{equation}
    \label{eqn:mom_thick}
    \delta_\theta = \int^{+\infty}_{-\infty} \left[\frac{u(y)}{u_\infty}\right] \left(1-\frac{u(y)}{u_\infty}\right) dy.
\end{equation}
The v-gutter's height ($L$) is kept constant for all the cases under consideration. The duct height is also kept at a constant value of $[h/L]=5.6$ (Figure \ref{fig:schematics}). Hence, a blockage value of about 18\% is expected for all the cases. Due to the wake intensity for different $\theta$ and $\xi$, $\delta_\theta$ varies appropriately. With increment in $\theta$ (Figure \ref{fig:flux_angle_slit}-IIIa), the upstream pressure gradient changes and increases $\delta_\theta$. It is also an indication that more momentum is lost to the bounding walls. Hence, the starting values of $\delta_\theta$ are less for lower $\theta$. In fact, $\delta_\theta$ drops almost to half of its value as $\theta$ drops from $[\pi/2]$ to $[\pi/6]$. As the v-gutter gradually accelerates the flow along the shoulder, values of $u$ increase. It is reflected as a drop-in $\delta_\theta$ values. Since the acceleration is the highest for $\theta = [\pi/6]$, the negative drop-in $\delta_\theta$ is also significant by almost 66.67\% in comparison with $\theta=[\pi/2]$. It has to be noted that the negative values in $\delta_\theta$ are a direct indicator of flow acceleration. The earlier drop present upstream the origin for $\theta = [\pi/6]$ is due to the placement of the v-gutter's nose-tip well ahead.

\begin{figure*}
	\centering{\includegraphics[width=\textwidth]{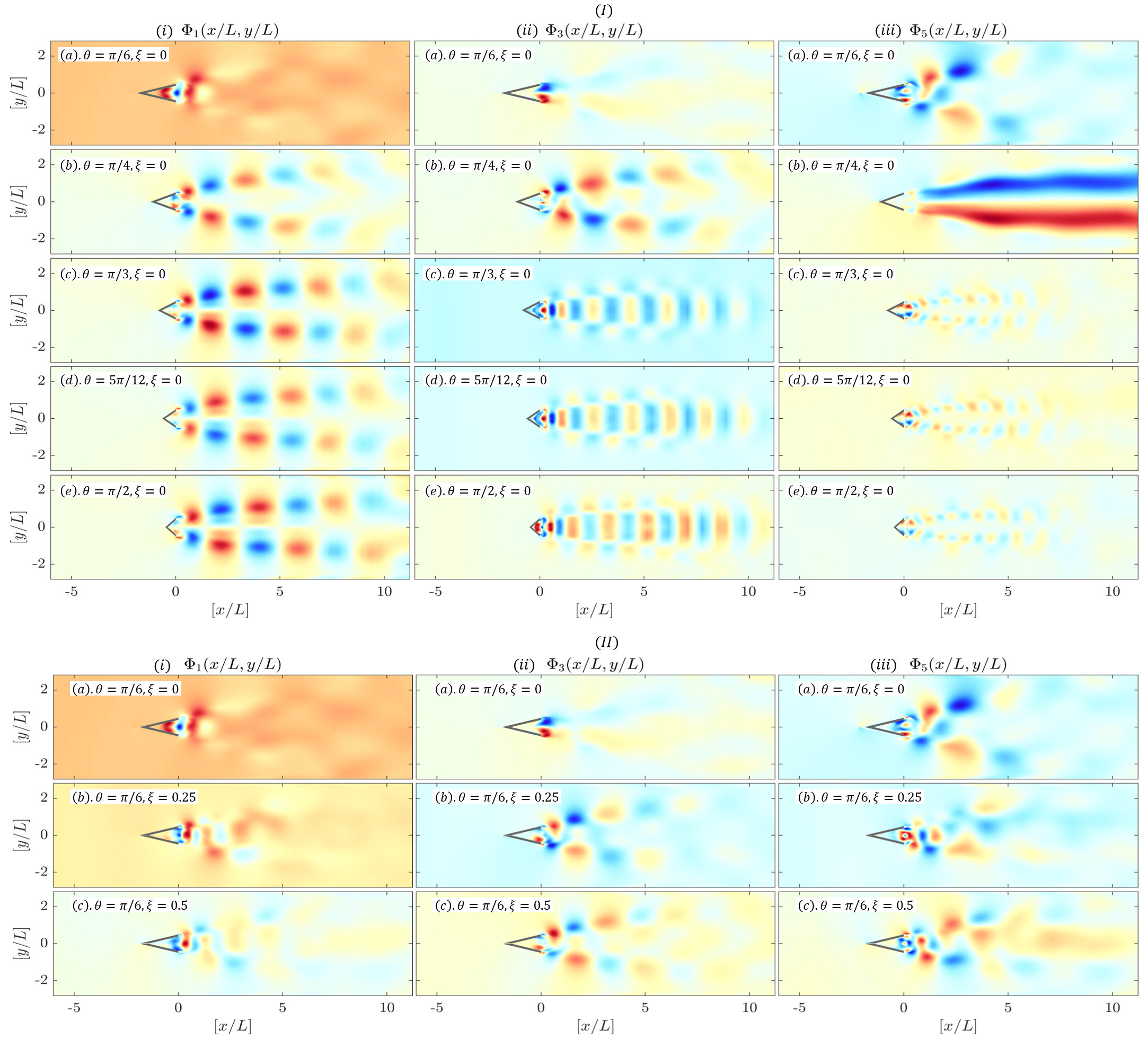}}
	\caption{Comparison of normalized energetic spatial mode obtained from the POD analysis of the velocity field ($\Phi_1,_3,_5[x/L,y/L]$) for v-gutter's different geometrical configurations. (I) Variation of v-gutter's angle in degree: (a). $\theta=[\pi/6],\xi=0$ mm, (b). $\theta=[\pi/4],\xi=0$ mm, (c). $\theta=[\pi/3],\xi=0$ mm, (d). $\theta=[5\pi/12],\xi=0$ mm, and (e). $\theta=[\pi/2],\xi=0$ mm. (II) Variation of v-gutter's slit size in mm: (a). $\theta=[\pi/6],\xi=0$ mm, (b). $\theta=[\pi/6],\xi=0.25$ mm, and (c). $\theta=[\pi/6],\xi=0.5$ mm. (Note: the red and blue color contour correspond to a value of +1 and -1 which can be interpreted as the positive and negative spatial correlation)}
	\label{fig:spatial_mode}
\end{figure*}

Downstream the v-gutter, due to the low recirculation region in $\theta = [\pi/6]$, wake flow quickly recovers to the freestream direction, as explained by the streamlines in the previous paragraphs. Hence the values of $\delta_\theta$ quickly recover closer to zero; however, it remains negative due to the accelerated flow's presence. For higher cases of $\theta$, the recirculation bubble elongates, leading to the formation of wavy streamlines. It is, in turn, reflected in the $\delta_\theta$ variation along the downstream direction. Also, the higher $\theta$, the streamwise velocity accelerates comparatively less, resulting in a smaller negative drop-in $\delta_\theta$ closer to the shoulder tip. For the cases of $\xi$ variation (Figure \ref{fig:flux_angle_slit}-IIIb), $\delta_\theta$ variation is small in the upstream. The presence of slit offers less adverse pressure gradient in contrast to the $\theta$ cases. Hence the peak drop-in $\delta_\theta$ is only 33.33\% between $\xi = 0$ and 0.5. Similarly, the wavy streamlines present downstream the v-gutter, with an almost periodic vortex shedding, explain the deviations in the downstream $\delta_\theta$ variations.

\subsection{Spatiotemporal modes} \label{ssec:spatiotemporal}

 In the current study, the authors have shown that a set of aperiodic and periodic cases exists as one changes the v-gutter's $\theta$. Moreover, it is also shown that the periodicity is also influenced by the slit size ($\xi$) for a given $\theta$. Wake-dominated flows exhibit dominant spatiotemporal characteristics. Performing modal analysis\cite{Taira2017,Kutz2016} like the Proper Orthogonal Decomposition\cite{Meyer2007} (POD) and Dynamic Mode Decomposition\cite{Schmid2010} (DMD) reveal dominant spatiotemporal modes, in terms of flow energy and the intrinsic dynamics. Moreover, such an analysis helps to see the prominent or difference in the spatiotemporal variations for the cases of $\theta$ and $\xi$. The underlying principle of both types of modal analysis is discussed elaborately in the open literature\cite{Bergmann2005,Janocha2022,Gao2021,Kumar2019,Yin2020,Du2022}. Hence, further discussion in the current manuscript is avoided. The dominant energetic and dynamic spatiotemporal modes from the modal analysis are shown in Figure \ref{fig:temporal_mode} and Figure \ref{fig:spatial_mode}.

In the analysis, one of the flow variables, such as the velocity magnitude at a given time step, is converted into a column matrix as mentioned in the processing of CFD data by Karthick\cite{Karthick2021}. Likewise, a two-dimensional matrix is prepared by stacking the column matrix for the entire solution time (i.e., 50 ms at a sampling rate of 20 kHz, thereby stacking 1000 snapshots). The obtained snapshot matrix is subjected to covariance matrix transformation and successive eigenmode decomposition in the POD analysis. The energy contained in each of the modes is ranked from the highest to the lowest, along with the dominant spatial structures are one of the critical outcomes of the POD analysis. Single value decomposition and further algebraic operations are done on the snapshot matrix in DMD analysis. The overall dynamic response obtained in terms of amplitude and frequency is one of the primary outcomes of the DMD analysis.

In Figure \ref{fig:temporal_mode} a-b, cumulative energy contents of the ranked modes \cite{Sahoo2021,Nanda2021} are shown for the variations in $\theta$ and $\xi$. The dominant mode ($k=1$) contains roughly $~20\%$ of the total energy in the fluid domain in $\theta$ variation cases (Figure \ref{fig:temporal_mode}a). It remains almost around the same value for all the cases of $\theta$ variations except with small increments as $\theta$ varies. However, in the cases of $\xi$ variations (Figure \ref{fig:temporal_mode}b), a drop closer to $~10\%$ is observed for $\xi=0.5$. The drop in the total energy in the first mode indicates that the energy is distributed to other new modes that drive the flow. It can be seen by tracking the cumulative energy variation with increasing mode numbers. It requires about six modes to represent the flow field with 50\% of energy in the plain case ($\xi=0$) than the slitted case ($\xi>0$), which requires about ten modes to represent the same. A higher number of modes indicate the presence of a multitude of flow dominant features. On the other hand, $\theta$ variation cases exhibit differences in the mode requirement to represent 50\% of the flow energy based on the aperiodic and periodic nature. Aperiodic cases require about 4-5 modes ($\theta = \pi/6$ and $\pi/4$), whereas periodic cases ($\theta = \pi/3, 5\pi/12,$ and $\pi/2$) require only three modes to represent the same level. Thus, the periodic cases must have a dominant flow structure encountered in the flow.

In Figure \ref{fig:temporal_mode} c-d, the dynamic modes\cite{Sahoo2020,Sugarno2022} are shown for both $\theta$ and $\xi$ variations. The dominant spectral content in the flow field is identified to drive the flow fluctuations in the domain. The non-dimensional spectra in Figure \ref{fig:temporal_mode}c reveal the presence of low amplitude, broadened spectra for the aperiodic case ($\theta = \pi/6$). For $\theta = [\pi/4]$, a discrete frequency is seen, however, with a large bandwidth representing the presence of both periodic and aperiodic features in the flow. Discrete, narrow bandwidth spectra are seen for all the other $\theta$ cases. The spectra seem to be decreasing gradually, with the non-dimensional spectra closely following the trend seen in Figure \ref{fig:spectra_angle_slits}. A series of tones can also be seen with a fundamental around $[fl/u_\infty]\sim 0.2$ and harmonics at $n = 2, 3,$ and 4. Although a trend in spectral decay is seen for the periodic cases, the amplitudes are not monotonous. It has to be noted that the deviations are not that significant. The wake structure formed in each case might be slightly different due to how the field is discretized during the post-processing. While looking at Figure \ref{fig:temporal_mode}d, the spectra for $\xi$ variations are seen. Broadened spectra are seen in the plain case ($\xi=0$), whereas a widened discrete spectrum starts to form for increasing $\xi$. Thus, the dynamic modes suggest the presence of periodic shedding with combined modes coming from the distorted recirculation bubble, as explained in the previous section.

In Figure \ref{fig:spatial_mode}, the dominant energetic spatial modes are listed for the variations of $\theta$ and $\xi$. Only the dominant three modes are discussed as it contains at-least 25\% of the total flow energy. Some of the odd leading modes look like the even leading modes as they are merely a pair\cite{Rao2019,Rao2020} due to the convection of dominant structures involved in the flow problem. For the case of $\theta=[\pi/6]$ (Figure \ref{fig:spatial_mode}-I i-a), the first dominant energetic spatial mode ($\Phi_1$) is globally chaotic with no significant structures visible except a few alternate structures in the wake. The chaotic behavior in the spatial field indicates that the flow is aperiodic with no definite structures. However, for the second dominant mode ($\Phi_3$), symmetric shedding vortices are seen closer to the v-gutter's wake. Similarly, for $\Phi_5$, shedding structures along the v-gutter's shoulder are seen with opposing coherence strength. Such a pattern is an indicator of asymmetric vortex shedding in the wake.

In the case of $\theta=[\pi/4], $ $\Phi_1$ and $\Phi_3$ are observed to contain periodic shedding structures along the v-gutter's shoulder tip with a wide divergence angle seen between the top and bottom structures. The third dominant mode ($\Phi_5$) is a shear mode as both the top and bottom shear layer away from the near wake zone are identified as coherent zones. In cases of $\theta=[\pi/3],[5\pi/12],$ and $[\pi/2]$, the first dominant structure ($\Phi_1$) is a periodic asymmetric shedding one and the second mode ($\Phi_3$) is a periodic symmetric shedding. Higher modes like $\Phi_5$ contain structures of lower wavelength and higher frequencies whose energy contents are only minimal, as described in Figure \ref{fig:temporal_mode}a. Thus, the behavior of periodic shedding for $\theta=[\pi/3]$ to $[\pi/2]$ can be corroborated by the presence of dominant periodic asymmetric shedding modes seen from the POD analysis.

Interesting spatial modes are seen for the cases of $\xi$ variations (Figure \ref{fig:spatial_mode})-II. As the slit size increases from $\xi = 0.25$ to $0.5$, the first dominant spatial modes $\Phi_1$ remain chaotic with localized structures, as seen in the plain case described a few paragraphs before. Events appear different in the second dominant spatial mode ($\Phi_3$). Instead of a near-wake symmetric structure representing the recirculation region's dominance, periodic asymmetric shedding structures are seen. The coherence strength is pronounced much for $\xi=0.5$ than $0.25$ as the slit balances the pressure in the wake and introduces an almost periodic flow. Such behavior in the dominant modes is the reason for the gross observation of almost periodic flow in the slitted case. All the higher modes like $\Phi_5$ are merely chaotic with small-scale structures with the presence of locally organized coherent structures. However, as discussed before in Figure \ref{fig:temporal_mode}b, most of these higher-order modes contain equivalent energy contents as that of the dominant ones. Hence, to represent the flow faithfully, they should also be considered.

\section{Conclusion} \label{sec:conclusion}
In the present manuscript, two-dimensional detached eddy simulation (DES) is performed on a bluff body flow problem at a compressible flow Mach number and Reynolds number of 0.25 and $0.1 \times 10^6$, respectively. V-gutter is chosen as the bluff body under investigation due to its potential use in combustors for flame holding and stabilization. Geometrical feature like the v-gutter's divergence angle is varied as $\theta = [\pi/6, \pi/4, \pi/3, 5\pi/12,$ and $\pi/2]$ in degree. In addition, the effect of introducing a slit to the v-gutter's stagnation zone is also considered. The size of the slit is varied as $\xi=0, 0.25$ and $0.5$ in mm, and the gross influence is reported only for $\theta = [\pi/6]$ to emphasize its influence on general wake flow dynamics. Following is a summary of some of the significant conclusions from the present study:
\begin{itemize}
    \item {Instantaneous and time-averaged flow studies reveal the presence of aperiodic vortex shedding behind the v-gutter for $\theta = [\pi/6]$. An almost periodic shedding is seen for $\theta = [\pi/4]$. For the cases of $\theta = [\pi/3, 5\pi/12$ and $\pi/2]$, periodic vortex shedding are seen.}
    \item {When the v-gutter is provided with a slit for $\theta = [\pi/6]$, the aperiodic flow pattern begins to transform to almost a periodic shedding. The intensity of periodic vortex shedding increases with $\xi$ from 0.25 to 0.5.}
    \item{On a general note, the alteration of streamwise velocity that is responsible for the generation of vorticity and subsequent shedding varies significantly with $\theta$ and $\xi$ near the shoulder tip. The perturbed recirculation bubble and the resulting global instability are attributed to the observations seen so far.}
    \item{The streamlined ($\theta=\pi/6$) and bluff body ($\theta=\pi/2$) case among the considered v-gutter geometries introduce varying total pressure loss ($\Delta p_0$). A maximum increment of 140\% in $\Delta p_0$ is seen between the two extreme cases due to the larger wake zone offered by the bluff body. When the slit is introduced, no noticeable increment is seen in $\Delta p_0$ asides from minor variations.}
    \item{In terms of drag coefficient variation ($c_d$), the effect of increments in $\theta$ reflects almost linearly in $c_d$. A maximum rise of $\sim 63\%$ is seen between the streamlined and bluff body. Similarly, increment increment in $\xi$ is observed to immediately decrease $c_d$ to a maximum of $\sim14\%$.}
    \item {From the spectral analysis, the dominant aperiodic frequency ($fL/u_\infty \sim 0.08 $ for $\theta=\pi/6$) is seen to be comparatively smaller than the periodic case. However, with $theta$, the dominant frequency in case of the periodic shedding ($fL/u_\infty \sim 0.18$ for $\theta=\pi/2$) decreases. For the slitted cases ($\xi$), the dominant frequency is observed to increase compared to $\theta$.}
    \item{The upstream fluctuations in velocity ($u,v$) and pressure ($p$) drop drastically with increments in $\theta$, whereas the trend is opposite with increments in $\xi$. Fluctuations in $p$ alone drop by $\sim 47\%$ between the streamlined and bluff body.}
    \item{From the $x-t$ analysis, in the aperiodic cases, the shedding structures are seen to be convecting at a comparatively higher velocity ($u_c \sim 75$ m/s) than that of the periodic cases ($u_c \sim 60$ m/s).}
    \item{From the momentum deficit ($\delta_\theta$) analysis, the upstream $\delta_\theta$ is affected significantly and seen to be decreasing with $\theta$ but remains almost unaffected with $\xi$. The periodic shedding cases exhibit localized variations with -$\delta_\theta$ owing to the flow acceleration behind the v-gutter, and the behavior is also seen for the slitted cases ($\xi$).}
    \item{From the modal analysis of the velocity magnitude snapshots, dominant spatial structures are identified. For $\theta$ variations, aperiodic shedding cases contain chaotic spatial structures in the dominant mode, whereas the periodic ones contain asymmetric shedding structures. The second dominant mode of the $\xi$ cases also contains a similar structure which explains the almost periodic vortex shedding behavior at $\xi=0.25$ and 0.5 cases.}
\end{itemize}

\section*{Supplemental material}
There is no supplemental material given along with the manuscript asides from the multimedia view files embedded with the figures.

\section*{Authors' Contributions}
LKG, and SKK, have contributed equally to this paper as the first authors. SKK, ARS, and RK have contributed to framing the problem statement and discussing the results. All the authors have been involved during the analysis and writing phase. The authors report no conflicts of interest.

\section*{Acknowledgement}
 All the authors thank the reviewer for their time reviewing the manuscript. The first author thanks the department and the college for providing the necessary fund to complete his doctoral thesis. The second author would like to extend his sincere thanks to his post-doctoral supervisor Prof. J. Cohen to encourage him to participate in research activities at leisure independently.

\section*{Data availability statement}
The data supporting this study's findings are available from the corresponding author upon reasonable request.

\section*{References}

%\bibliographystyle{aipnum4-2}
\bibliography{references}

\onecolumngrid

\PRLsep	

\end{document}